\documentclass{jfm}
\usepackage{graphicx}
\usepackage{epstopdf, epsfig}
\usepackage{color}
\usepackage{amsmath}
\usepackage{amssymb}
\usepackage{natbib}
\usepackage{fancyhdr}
\usepackage[dvipsnames]{xcolor}
\usepackage[normalem]{ulem}

\newcommand\cites[1]{\citeauthor{#1}'s\ (\citeyear{#1})}

\DeclareMathOperator{\ImIm}{\mathrm{Im}}
\DeclareMathOperator{\ReRe}{\mathrm{Re}}

\shorttitle{MHD instability on a sphere}
\shortauthor{C. Wang, A.D. Gilbert and J. Mason}

\title{An analytical study of the MHD clamshell instability on a sphere}

\author{Chen Wang,
    \corresp{\email{C.Wang6@exeter.ac.uk}}
  Andrew D. Gilbert
 \and Joanne Mason}

\affiliation{Department of Mathematics and Statistics, \\[3pt]
University of Exeter,
Exeter,  EX4 4QF,  UK}

\begin{document}

\maketitle


\abstract
This paper studies the instability of two-dimensional magnetohydrodynamic (MHD) systems on a sphere using analytical methods.  The underlying flow consists of a zonal  differential rotation and a toroidal magnetic field is present. Semicircle rules that prescribe the possible domain of the wave velocity in the complex plane for general flow and field profiles are derived.
The paper then sets out an analytical study of the `clamshell instability', which features field lines on the two hemispheres tilting in opposite directions (Cally 2001, \textit{Sol. Phys.} vol. 199, pp. 231--249). An asymptotic solution for the instability problem is derived for the limit of weak shear of the zonal flow, via the method of matched asymptotic expansions. It is shown that when the zonal flow is solid body rotation, there exists a neutral mode that tilts the magnetic field lines,
 referred to as the `tilting mode'.  A weak shear of the zonal flow excites the critical layer of the tilting mode,
 which reverses the tilting direction to form the clamshell pattern and induces the instability.  The asymptotic solution provides
 insights into properties of the instability for a range of flow and field profiles.  A remarkable feature is that the magnetic field affects the instability only through its local behaviour in the critical layer.


\section{Introduction}


Magnetohydrodynamic (MHD) instability is of significant importance to astrophysical flows. Magnetic fields are ubiquitous in stars and planets, and although the magnetic field can act as a restoring force, it can also destabilise the fluid, resulting in turbulence and flow rearrangement. 

There are numerous MHD instabilities, differing in geometry and parameter regime. In this study, our particular focus is on two-dimensional MHD instability on a sphere. This has important applications to the solar tachocline, which is a thin transition layer between the Sun's radiative interior and the outer convection zone. The tachocline couples these two regions that have distinct properties, and it plays a pivotal role in solar physics.
In particular, this thin shear layer is believed to be the seat of the solar dynamo (see, for example, \cite{Charbonneau14,Brun17}).
Strong azimuthal magnetic field generated in the tachocline rises up to the photosphere via magnetic buoyancy, and generates a variety of surface phenomena including sunspots, coronal loops and flares. MHD instabilities in the solar tachocline can significantly modify the magnetic field and thus have a strong impact on the subsequent surface phenomena.


In this context,
there have been numerous studies of MHD instabilities of zonal flow in a thin spherical shell coupled with a toroidal magnetic field. 
The differential rotation profile of the Sun may be modelled as the angular velocity \citep{Newton51}
\begin{equation}\label{0.1}
  \Omega=r-s\mu^2,
\end{equation}
where $r$ is the angular velocity at the equator, $s$ is the shear rate and $\mu=\cos\theta$, $\theta$ being the colatitude in spherical polar coordinates. The parameters $r$ and $s$ are both positive for solar differential rotation, as the Sun rotates faster at the equator. For hydrodynamic flows without magnetic fields, \citet{Watson81}  found that the zonal flow  (\ref{0.1}) becomes unstable when $s/r>0.29$. For the Sun, $s/r=0.12\sim0.17$ (see, for example, \citet{Gough07}) and the flow is hydrodynamically stable. When a magnetic field is added, however, \citet{Gilman97} 
have shown that instabilities may be present for $s/r$ smaller than 0.29: these instabilities require  weaker shear, implying that the magnetic field has a destabilising effect. \citet{Gilman97}  named such instabilities `joint instabilities' since they only arise when both the hydrodynamic shear and the magnetic field, which are stable separately, exist together. Joint instabilities exist for a wide range of toroidal magnetic fields, including broad field profiles with single or multiple nodes \citep{Gilman97,Gilman99}, and magnetic field bands localised at various latitudes \citep{Dikpati99,Gilman00b}. 

The `singular points' of the unstable modes play an important role in  this joint instability. These points are located where the phase velocity of the mode relative to the basic zonal flow matches the characteristic velocity of the Alfv\'{e}n waves. Here the  wave equation becomes singular for ideal fluids; weak effects of viscosity or magnetic resistivity, unsteadiness or nonlinearity  remove the singularity, but the disturbances still exhibit  strong amplitudes locally.  In hydrodynamic stability theory, the singular points and their vicinity are named `critical levels' and `critical layers'  \citep{Drazin82}, terminology we will adopt here. 
\citet{Gilman99} and \citet{Dikpati99} have found that the disturbances change dramatically across the critical layers,  and that the critical layers largely determine the spatial structure of the unstable modes. Moreover, various stresses that contribute to the energy of the instability are concentrated in the critical layers, suggesting that they are responsible for driving the instability.

 \citet{Cally01} and \citet{Cally03} computed the  nonlinear evolution of the joint instability for two-dimensional flow on a sphere.  For strong and broad magnetic field profiles, the magnetic field lines feature a `clamshell' pattern in the early stage:  the field lines are tilted in opposite directions on the two hemispheres, and they named this the `clamshell instability'. In the later nonlinear evolution, the field lines tilt over $90^\circ$ and  reconnect  at the equator. These behaviours remain similar when the more realistic physics of density stratification and vertical shear are added, though they produce more subtle vertical structures  \citep{Miesch07}. If an external force that  maintains the zonal flow and the poloidal field is added \citep{Miesch07sustained},  then the field lines do not tilt over. Instead, the clamshell pattern is maintained together with strong mean toroidal field. 
The mean field and the unstable mode dominate in turn in a quasi-periodic manner.


In the present study, we undertake analytical investigations
to understand two-dimensional MHD instability on a sphere. We derive the semicircle rules that prescribe the domain of the complex phase velocities for general profiles of a zonal flow and a toroidal magnetic field. A semicircle rule was first derived by \citet{Howard61} for hydrodynamic instability in Cartesian geometry. It states that for any unstable mode in an inviscid parallel shear flow $U(y)$, the phase velocity $c=c_\mathrm{r}+\mathrm{i}c_\mathrm{i}$ must lie in a semicircle  with center $(c_\mathrm{r},c_\mathrm{i})=( \tfrac{1}{2}(U_\mathrm{max}+U_\mathrm{min}),0)$ and radius $\tfrac{1}{2}(U_\mathrm{max}-U_\mathrm{min})$ above the real axis. It has become a celebrated result of hydrodynamic instability for its generality and simple form. \citet{Watson81}, \citet{Thuburn96} and \citet{Sasaki12}  extended the theory to hydrodynamic instability in spherical geometry, showing that this introduces additional terms in the radius of the semicircle. On the other hand, \citet{Howard62}, \citet{Gilman67}, \citet{Chandra73}, \citet{Cally00}, \citet{Hughes01} and \citet{Deguchi21}  derived semicircle rules for MHD instability in Cartesian geometry. Their results indicate that the magnetic field can reduce the radii of the semicircles. Consequently, if the magnetic field is strong enough everywhere, the semicircles will disappear and stability is guaranteed. In this paper, we derive semicircle rules for MHD instability in spherical geometry. An interesting phenomenon is that in this geometry, the magnetic field can increase the radii of the semicircles, something that never happens in the Cartesian case.

We also develop an asymptotic analysis for the clamshell instability. The limit that we consider is when the shear of the zonal flow is relatively weak, a typical situation in solar physics: in equation (\ref{0.1}), $s/r$ is a small number for solar differential rotation. For the magnetic field, we consider the broad and strong field profiles which are typical for the clamshell instability \citep{Cally01}. The location where the magnetic  field profile passes through zero, the node of the profile, is where the magnetic critical level sits and this plays a fundamental role in the instability (e.g., see \citet{Gilman99}). Here, in our first study of the problem, we limit our attention to the situation where there is only one such node of the field profile and we assume that its gradient there is non-zero. We derive the solution to the instability problem for general profiles from this family.

Our asymptotic analysis can help us better understand the clamshell instability in several ways. First, it can provide an analytical explanation for the mechanism of the instability: we show that the instability is caused by the interaction between the global tilting motion of the magnetic field and the critical level of the mode located at the node of the field profile.  Second, it can give insights for general profiles of magnetic field and shear flows: which types of profiles are unstable and which are not. Finally, it can easily tackle the neutral stability limit where numerical solutions often suffer from resolution difficulties caused by singularities emerging in the corresponding eigenmodes.

The organisation of the paper is as follows. In \S 2  we present the governing equations of two-dimensional MHD flow on a sphere, and the corresponding equations for the linear instability problem.  The equations for the mean-flow response and angular momentum conservation are also given.  In \S 3  we derive the semicircle rules for the complex phase velocity and we discuss their applications.  In \S4  we undertake an asymptotic analysis for the clamshell instability.  In \S 4.1 we  show that the tilting mode exists for solid body rotation,  and that a weak shear can excite its critical levels.  Then we solve the eigenvalue problem in \S 4.2 by matching the tilting mode and critical layer. The results of the eigenvalue problem, which yield a number of general conclusions, are discussed in \S\ref{S5.1}--\ref{S5.4},  and the conservation of angular momentum is shown to provide a mechanism for the instability in \S \ref{S5.5}.  Concluding remarks are given in \S \ref{S6}. 

\section{Governing equations} \label{S2}

In this section we present the equations of two-dimensional MHD on a sphere and derive the equations governing the linear instability problem.
We consider the motion of an incompressible, inviscid, perfectly electrically conducting fluid with density $\rho$ and magnetic permeability $\mu$. The flow is on a sphere of radius $R$ with a characteristic velocity $U_0$. The MHD equations for the dimensionless velocity $\textit{\textbf{u}}$, magnetic field $\textit{\textbf{B}}$, and pressure $p$ are
\begin{equation}
	\nabla\cdot\textit{\textbf{u}} \label{1.1}
	=0,
\end{equation}
\begin{equation}
	\nabla\cdot\textit{\textbf{B}}=0,
\end{equation}
\begin{equation}
	\frac{\partial\textit{\textbf{u}}}{\partial t}+(\textit{\textbf{u}}\cdot\nabla) \textit{\textbf{u}}=-\nabla (p+\tfrac{1}{2} {B^2})+(\textit{\textbf{B}}\cdot  \nabla )\textit{\textbf{B}},
\end{equation}
\begin{equation}
	\frac{\partial \textit{\textbf{B}}}{\partial t}+(\textit{\textbf{u}}\cdot \nabla) \textit{\textbf{B}}= (\textit{\textbf{B}}\cdot \nabla )\textit{\textbf{u}}, \label{1.4}
\end{equation}
where the  length, time, velocity, magnetic field and pressure have been nondimensionalised by $R$, $R/U_0$, $U_0$, $U_0\sqrt{\mu\rho}$ and $\rho U_0^2$, respectively. We use spherical polar coordinates $(r,\theta,\phi)$, where $r$ is the radius, $\theta$ is the co-latitude and $\phi$ is the longitude, and the corresponding unit vectors are $(\textit{\textbf{e}}_r, \textit{\textbf{e}}_\theta, \textit{\textbf{e}}_\phi )$. We take the system  to be two-dimensional, as for a thin spherical shell, so that the velocity and magnetic field have no radial component (that is, in the $\textit{\textbf{e}}_r$-direction), and are independent of the radial coordinate $r$.  

On writing
$\textit{\textbf{u}}= v\textit{\textbf{e}}_\theta+ u\textit{\textbf{e}}_\phi$ and $\textit{\textbf{B}}= b\textit{\textbf{e}}_\theta + a\textit{\textbf{e}}_\phi$, where $u$, $v$, $a$ and $b$ are functions of $\theta$, $\phi$ and $t$,
equations (\ref{1.1})--(\ref{1.4}) become




\begin{equation}
\frac{\partial u}{\partial \phi}+\frac{\partial}{\partial \theta}(v\sin \theta)=0, \label{2.5}
\end{equation}
\begin{equation}
\frac{\partial a}{\partial \phi}+\frac{\partial}{\partial \theta}(b\sin \theta)=0,
\end{equation}
\begin{equation}
\frac{\partial u}{\partial t}+\frac{u}{\sin \theta}\frac{\partial u}{\partial \phi}+v\, \frac{\partial u}{\partial \theta}+\frac{uv\cos \theta}{\sin \theta}=-\frac{1}{\sin \theta}\frac{\partial p}{\partial\phi}-\frac{b}{\sin \theta}\frac{\partial b}{\partial \phi}+b\,\frac{\partial a}{\partial \theta}+\frac{ab \cos \theta}{\sin \theta},
\end{equation}
\begin{equation}
\frac{\partial v}{\partial t}+\frac{u}{\sin \theta}\frac{\partial v}{\partial \phi}+v\,\frac{\partial v}{\partial \theta}-\frac{u^2\cos \theta}{\sin \theta}=-\frac{\partial p}{\partial \theta}-a\,\frac{\partial a}{\partial \theta}+\frac{a}{\sin \theta}\frac{\partial b}{\partial \phi}-\frac{a^2\cos\theta}{\sin \theta},
\end{equation}
\begin{equation}
\frac{\partial a}{\partial t}-\frac{\partial}{\partial \theta}(ub-va)=0, \label{2.10}
\end{equation}
\begin{equation}
\frac{\partial b}{\partial t}+\frac{1}{\sin \theta}\frac{\partial}{\partial \phi}(ub-va)=0. \label{2.11}
\end{equation}
%
We consider an axisymmetric basic state consisting of a zonal flow and a toroidal magnetic field: $u=U$, $v=0$, $a=A$, $b=0$, where $U$ and $A$ vary with co-latitude $\theta$ but are independent of longitude $\phi$. The basic state pressure $P$ is therefore governed by the balance
\begin{equation}
	\frac{\partial}{\partial \theta}(P+\tfrac{1}{2}{A^2})= (U^2-A^2)\cot \theta.
\end{equation}
We then study the instability of this state to small disturbances:
\begin{equation}
	u=U+u_\ell,\quad v=v_\ell,\quad a=A+a_\ell,\quad b=b_\ell, \quad p=P+p_\ell, \label{2.12}
\end{equation}
where the  $\ell$ subscript denotes disturbances of linear instability. Substituting (\ref{2.12}) into (\ref{2.5})-(\ref{2.11}) and linearising gives
\begin{equation}
\frac{\partial u_\ell}{\partial \phi}+\frac{\partial}{\partial \theta}(v_\ell\sin \theta)=0, \label{1.15}
\end{equation}
\begin{equation}
\frac{\partial a_\ell}{\partial \phi}+\frac{\partial}{\partial \theta}(b_\ell\sin \theta)=0, \label{1.16}
\end{equation}
\begin{equation}
	\frac{\partial u_\ell}{\partial t}+\frac{U}{\sin \theta}\frac{\partial u_\ell}{\partial \phi}+\frac{\mathrm{d}U}{\mathrm{d}\theta}\, v_\ell+\frac{U\cos\theta}{\sin \theta}\, v_\ell=-\frac{1}{\sin\theta}\frac{\partial p_\ell}{\partial \phi}+\frac{\mathrm{d}A}{\mathrm{d}\theta}\, b_\ell+\frac{A\cos\theta}{\sin \theta}\, b_\ell, \label{2.15}
\end{equation}
\begin{equation}
	\frac{\partial v_\ell}{\partial t}+\frac{U}{\sin\theta}\frac{\partial v_\ell}{\partial \phi}-\frac{2U\cos \theta}{\sin \theta}\, u_\ell=-\frac{\partial}{\partial\theta}\left(p_\ell+Aa_\ell\right)+\frac{A}{\sin\theta}\frac{\partial b_\ell}{\partial\phi}-\frac{2A\cos\theta}{\sin\theta}\, a_\ell, \label{2.16}
\end{equation}
\begin{equation}
	\frac{\partial a_\ell}{\partial t}-\frac{\partial}{\partial \theta}\left(Ub_\ell-Av_\ell\right)=0, \label{1.19}
\end{equation}
\begin{equation}
	\frac{\partial b_\ell}{\partial t}+\frac{1}{\sin\theta}\frac{\partial}{\partial \phi}(Ub_\ell-Av_\ell)=0. \label{1.20}
\end{equation}

For mathematical convenience, we then introduce the notation
\begin{equation}
U(\theta)=\Omega(\theta)  \sin\theta,\quad A(\theta) =\beta(\theta) \sin\theta\,.
\end{equation}
With the radius of the sphere as unity in our non-dimensional system, $\Omega$ is the 
angular velocity, and $\beta$ is the magnetic analogue as noted by \citet{Gilman97}. For a flow and magnetic field that are smooth at the poles $\theta=0$, $\pi$, the quantities $\Omega$ and $\beta$ will tend to constants there. In the absence of a better term, we will refer to $\beta$, inaccurately, as the magnetic \emph{field} from now on.
In view of the divergence-free conditions (\ref{1.15}) and (\ref{1.16}), we introduce the stream function $\psi$ and the flux function $\chi$, such that
\begin{equation}
	u_\ell=-\frac{\partial \psi}{\partial \theta}\, ,\quad v_\ell=\frac{1}{\sin\theta}\frac{\partial \psi}{\partial \phi}\, ,\quad a_\ell=-\frac{\partial\chi}{\partial \theta}\, ,\quad b_\ell=\frac{1}{\sin\theta}\frac{\partial\chi}{\partial\phi}\, . \label{2.20}
\end{equation}
We combine (\ref{2.15}) and (\ref{2.16}) to eliminate the pressure $p$, and then we apply  (\ref{2.20}). After some algebra, following \citet{Watson81}, we derive the vorticity equation
\begin{align}
\left(\frac{\partial}{\partial t}+\Omega\frac{\partial}{\partial \phi}\right)&\nabla^2\psi-\frac{1}{\sin \theta}\frac{\mathrm{d}}{\mathrm{d}\theta}\left[\frac{1}{\sin\theta}\frac{\mathrm{d}}{\mathrm{d}\theta}(\Omega\sin^2\theta)\right]\frac{\partial \psi}{\partial \phi} \nonumber \\
-\beta\frac{\partial}{\partial \phi}&\nabla^2\chi+\frac{1}{\sin \theta}\frac{\mathrm{d}}{\mathrm{d}\theta}\left[\frac{1}{\sin\theta}\frac{\mathrm{d}}{\mathrm{d}\theta}(\beta\sin^2\theta)\right]\frac{\partial \chi}{\partial \phi}=0, \label{1.24}
\end{align}
where the Laplacian on the spherical surface is
\begin{equation}
	\nabla^2 f \equiv\frac{1}{\sin\theta}\frac{\partial }{\partial\theta}\left(\sin\theta\, \frac{\partial f}{\partial\theta}\right)+\frac{1}{\sin^2\theta}\frac{\partial^2 f}{\partial\phi^2}\, .
\end{equation}
Similarly, using (\ref{2.20}) in the induction equation, (\ref{1.19}) and (\ref{1.20}), we obtain
\begin{equation}
\left(\frac{\partial}{\partial t}+\Omega\, \frac{\partial}{\partial \phi}\right)\chi-\beta\, \frac{\partial \psi}{\partial \phi}=0. \label{1.26}
\end{equation}

Now we consider normal mode disturbances, replacing
\begin{equation}\label{2.24a}
  (\psi, \chi)\rightarrow(\psi, \chi) \, e^{\mathrm{i}m(\phi-ct)}\,,
\end{equation}
where $m$ is an integer representing the wavenumber in the longitudinal direction and $c$ is a complex constant representing the phase velocity. Using the substitution $\mu=\cos\theta$,  (\ref{1.24}) and (\ref{1.26}) become ordinary differential equations in $\mu$:
\begin{equation}
	(\Omega-c)L\psi-\psi\, \frac{\mathrm{d}^2}{\mathrm{d}\mu^2}\bigl[\Omega(1-\mu^2)\bigr]-\beta L\chi+\chi\, \frac{\mathrm{d}^2}{\mathrm{d}\mu^2}\bigl[\beta(1-\mu^2)\bigr]=0,\label{1.27}
\end{equation}
\begin{equation}
(\Omega-c)\chi-\beta\psi=0,  \label{2.24}
\end{equation}
where $L$ is the Legendre operator
\begin{equation}
Lf \equiv \frac{\mathrm{d}}{\mathrm{d}\mu}\left[(1-\mu^2)\, \frac{\mathrm{d}f}{\mathrm{d}\mu}\right]-\frac{m^2 f}{1-\mu^2}\, . \label{2.25}
\end{equation}
We note that equations (\ref{1.27}) and (\ref{2.24}) do not hold for axisymmetric disturbances, having $m=0$ (and indeed are written down having been divided throughout by $\mathrm{i}m$). The continuity equation (\ref{1.15}) does not allow axisymmetric disturbances that remain finite at the poles and such disturbances are only possible 
when a free surface is present; see \citet{Gilman02}. Without loss of generality, we take $m$ to be a positive integer. 

Finally, for the subsequent analysis it is convenient to introduce the variable $H$ defined by
\begin{equation}\label{2.26}
  H=\frac{\psi}{\Omega-c}=\frac{\chi}{\beta}\, ,
\end{equation}
motivated by (\ref{2.24}).  Its governing equation is
\begin{equation}\label{2.27}
  (SH')'+\left[2(\Omega-c)(\mu\Omega)'-2\beta(\mu\beta)'-\frac{m^2S}{(1-\mu^2)^2}\right]H=0,
\end{equation}
with
\begin{equation}
  S=\bigl[(\Omega-c)^2-\beta^2\bigr](1-\mu^2), 
\end{equation}
where the prime denotes a derivative with respect to $\mu$. The physical meaning of $H$ is that it is the Lagrangian displacement in the $\theta$-direction scaled by $\sin\theta$. 

The boundary conditions for (\ref{1.27}), (\ref{2.24}) and (\ref{2.27})  are that  $\psi$, $\chi$ and $H$ should remain finite at the poles $\mu=\pm 1$, these locations being singularities of the  equations. This poses an eigenvalue problem for the phase velocity $c$. Our particular interest is the situation where we have a complex eigenvalue $c=c_\mathrm{r}+\mathrm{i}c_\mathrm{i}$ with $c_\mathrm{i}>0$,  indicating the presence of instability. In any case since equation (\ref{2.27}) is real, solutions for $c$ always appear in complex conjugate pairs.

Equation (\ref{2.27}) has singularities at locations $\mu_\star$ where $S=0$ or
\begin{equation}\label{2.30a}
  \Omega-c=\pm \beta.
\end{equation}
In general, $H$ diverges at such points if the eigenvalue $c$ is real, and $\mu_\star$ is referred to as a critical level.  In the context of instability, i.e. when  $c=c_\mathrm{r}+\mathrm{i}c_\mathrm{i}$, $c_\mathrm{i}>0$, the solution remains analytic (for real $\mu$) but there are strong gradients of the eigenfunctions in the critical layer where $\Omega-c_\mathrm{r}\approx \pm \beta$, since $c_\mathrm{i}$ is usually small. We will show that the critical layer plays a fundamental role in the instability problem.

A main contribution of this paper is an asymptotic analysis of the eigenvalue problem for the `clamshell instability', presented in \S \ref{asymptotics} in detail. In the limit of weak shear of the zonal flow $\Omega'$, we derive an asymptotic solution for the eigenvalue and eigenfunction using the method of matched asymptotic expansions, combining solutions in the bulk of the flow and the critical layer. We will also solve the problem numerically by adopting two methods from previous studies: one expands $\psi$ and $\chi$ using Legendre polynomials \citep{Gilman99}, and the other is a shooting method \citep{Dikpati99}.  The method of Legendre polynomial expansion computes all of the eigenvalues but it can be expensive. Similarly to \citet{Gilman99}, we use this method when $\Omega(\mu)$ and $\beta(\mu)$ are expressed by polynomials, so that their expansions merely involve several terms, resulting in a sparse matrix for the eigenvalue problem. The shooting method is fast and provides more precise solutions, but it needs a good guess for the eigenvalue. Such a good guess will either come from the method of Legendre polynomial expansion or the asymptotic solution derived in the limit of weak shear. When these two approaches are not available, some trial and error for the initial guess has to be performed.


In what follows, we will also be interested in the mean-flow response of the linear instability.  We therefore set
\begin{equation}
	u=U+u_\ell+\Delta U,\quad v=v_\ell+\Delta V,\quad a=A+a_\ell+\Delta A,\quad b=b_\ell+\Delta B, \label{2.30}
\end{equation}
where the mean-flow modifications $\Delta U$, $\Delta V$, $\Delta A$ and $\Delta B$ are forced by the linear disturbances and are independent of $\phi$. Substituting (\ref{2.30}) into (\ref{2.5})--(\ref{2.11}),
applying the zonal average
 \begin{equation}
   \overline{f}(\theta) =\frac{1}{2\pi}\int_0^{2\pi} f (\theta, \phi)\, \mathrm{d}\phi, \label{2.31}
 \end{equation}
and noting the spatial periodicity of disturbances in $\phi$, we find that $\Delta V$ and $\Delta B$ are zero and $\Delta U$ and $\Delta A$ are governed by
\begin{equation}
  \frac{\partial \Delta U}{\partial t}=\frac{1}{\sin^2 \theta}\frac{\partial}{\partial \theta}\, \overline{\sin^2\theta (a_\ell b_\ell-u_\ell v_\ell)}\,,\label{2.32}
\end{equation}
\begin{equation}
  \frac{\partial\Delta A}{\partial t}=\frac{\partial}{\partial \theta}\, \overline{(u_\ell b_\ell -v_\ell a_\ell)}\, . \label{2.32b}
\end{equation}
In the spherical system, angular momentum and mean toroidal field are conserved, namely
\begin{equation}\label{2.33}
  \int_0^\pi 2\pi \sin^2\theta\, \frac{\partial \Delta U}{\partial t}\,\mathrm{d}\theta=0,
\end{equation}
\begin{equation}\label{2.34}
  \int_0^\pi  \frac{\partial\Delta A}{\partial t}\,\mathrm{d}\theta=0\,,
\end{equation}
as is guaranteed by (\ref{2.32}) and (\ref{2.32b}).

\section{The semicircle rules}


In this section, we derive  semicircle rules that provide general bounds for the eigenvalue, in the style of the celebrated theory of \citet{Howard61}. To tackle the spherical geometry, we will mainly follow the theory of \citet{Watson81}  who derived semicircle rules for hydrodynamic instability on a sphere. We will also provide alternative bounds to his theory which could be tighter for certain types of flows.

We proceed by multiplying equation (\ref{2.27}) by the complex conjugate $H^*$ of $H$ and integrating from $\mu=-1$ to $\mu=1$. Applying integration by parts, we derive the integral formula
\begin{eqnarray}
\int_{-1}^1\left[(\Omega-c)^2-\beta^2\right]\Bigl[(1-\mu^2)|H'|^2+\frac{m^2}{1-\mu^2}|H|^2\Bigr]\mathrm{d}\mu \nonumber\\
=\int_{-1}^1 2\bigl[(\Omega-c)(\mu\Omega)'-\beta(\mu\beta )'\bigr]|H|^2\, \mathrm{d}\mu. \label{5}
\end{eqnarray}
Writing the complex phase velocity as $c=c_{\mathrm{r}}+\mathrm{i}c_\mathrm{i}$ and taking $c_\mathrm{i}>0$ for an unstable mode,  the imaginary part of (\ref{5}) yields
\begin{equation}
\int_{-1}^1\Omega G\, \mathrm{d}\mu=\int_{-1}^1 \bigl[c_\mathrm{r} G+(\mu\Omega)'|H|^2\, \bigr]\, \mathrm{d}\mu \label{8}
\end{equation}
with
\begin{equation}
G=(1-\mu^2)\left|H'\right|^2+\frac{m^2}{1-\mu^2}\, |H|^2\ge 0. \label{7}
\end{equation}
This result has been derived by \citet{Gilman97}. The real part of (\ref{5}) is
\begin{equation}
\int_{-1}^1 (\Omega^2-2\Omega c_\mathrm{r}+c_\mathrm{r}^2-c_\mathrm{i}^2-\beta^2)G\, \mathrm{d}\mu=\int_{-1}^1 2 \bigl[(\Omega-c_\mathrm{r}) (\mu\Omega)'-\beta(\mu\beta)'\bigr]|H|^2 \, \mathrm{d}\mu. \label{9}
\end{equation}
Using (\ref{8}) to replace the second term on the left-hand-side of (\ref{9}),  we have
\begin{equation}
(c_\mathrm{r}^2+c_\mathrm{i}^2)\int_{-1}^1G\, \mathrm{d}\mu=\int_{-1}^1(\Omega^2-\beta^2)G\, \mathrm{d}\mu+ \int_{-1}^1 2\bigl[\beta(\mu\beta)'-\Omega(\mu\Omega)'\bigr]|H|^2\, \mathrm{d}\mu. \label{10}
\end{equation}
Now, following \citet{Howard61} and \citet{Watson81}, we quote the inequality
$$
\int_{-1}^1(\Omega-\Omega_\mathrm{max})(\Omega-\Omega_\mathrm{min})G\, \mathrm{d}\mu\leq 0,
$$
where $\Omega_\mathrm{max}$ and $\Omega_\mathrm{min}$ are the maximum and minimum values of $\Omega$ for all $\mu$, which gives
\begin{equation}
\int_{-1}^1\Omega^2 G\, \mathrm{d}\mu\leq \int_{-1}^1\bigl[\Omega(\Omega_{\max}+\Omega_\mathrm{min})-\Omega_\mathrm{max}\Omega_\mathrm{min}\bigr] G\, \mathrm{d}\mu. \label{11}
\end{equation}
Substituting (\ref{11}) into (\ref{10}) to replace the $\Omega^2G$ term, we derive
\begin{align}
(c_\mathrm{r}^2+c_\mathrm{i}^2)\int_{-1}^1G\, \mathrm{d}\mu&\leq \int_{-1}^1 \bigl[(\Omega_\mathrm{max}+\Omega_\mathrm{min})\Omega-\Omega_\mathrm{\max}\Omega_\mathrm{min}-\beta^2 \bigr]G\,  \mathrm{d}\mu \nonumber\\
&+ \int_{-1}^1  2 \bigl [\beta(\mu\beta)'-\Omega(\mu\Omega)'\bigr]|H|^2\mathrm{d}\mu.\label{12}
\end{align}
Finally, applying (\ref{8}) to (\ref{12}) again leads to the inequality
\begin{align}
\bigl[\left(c_\mathrm{r}-\overline{\Omega}\right)^2+c_\mathrm{i}^2\bigr]\int_{-1}^1G\, \mathrm{d}\mu
&\leq\int_{-1}^1\left(\Delta \Omega^2-\beta^2\right)G\, \mathrm{d}\mu \nonumber \\
&+\int_{-1}^1 2 \bigl[ \beta(\mu\beta)'-(\Omega-\overline{\Omega})(\mu \Omega)' \bigr]|H|^2\, \mathrm{d}\mu\,, \label{13}
\end{align}
where
\begin{equation}
  \overline{\Omega}=\frac{\Omega_\mathrm{max}+\Omega_\mathrm{min}}{2}\, ,\quad \Delta \Omega=\frac{\Omega_\mathrm{max}-\Omega_\mathrm{min}}{2}\, .
\end{equation}

We have obtained two relations for $c_\mathrm{r}^2+c_\mathrm{i}^2$ and $(c_\mathrm{r}-\overline{\Omega})^2+c_\mathrm{i}^2$, namely (\ref{10}) and (\ref{13}). In the case of MHD instability in Cartesian geometry, e.g. the study of \citet{Hughes01},  two similar equations hold but the integrals involving $|H|^2$ are not present. In that case, one can derive two semicircle rules straightforwardly by bounding the integrals of $G$. In our case of  spherical geometry, however, we need to consider how to bound the two $|H|^2$ integrals in order to find semicircle rules.
First, we note that we have one bound for $|H|^2$ from the definition of $G$ in (\ref{7}), namely
\begin{equation}
0\le |H|^2\le \frac{1-\mu^2}{m^2}\, G\,. \label{3.1}
\end{equation}
This holds at each location of $\mu$, and so we refer to it as the \emph{pointwise bound}. An alternative is to bound the integral of $|H|^2$. For this task, we invoke the theorem of Rayleigh's quotient. Let $L$ be a linear Sturm--Liouville operator and  $\lambda$ be the smallest eigenvalue for the corresponding Sturm--Liouville problem $L H+\lambda H=0$, with homogeneous boundary conditions at $\mu=a$ and $\mu=b$. Then, for arbitrary smooth functions $H(\mu)$, the Rayleigh quotient $R$ satisfies
\begin{equation}
	R=-\, \frac{\displaystyle\int_{a}^b H^*L H\, \mathrm{d}\mu}{\displaystyle\int_a^b|H|^2\, \mathrm{d}\mu}\ge \lambda. \label{3.11}
\end{equation}
The smallest eigenvalue of the Legendre operator $L$ defined in (\ref{2.25}) is $\lambda=m(m+1)$. Thus, applying integration by parts to the numerator of (\ref{3.11}), we obtain
\begin{equation}
0\le \int_{-1}^1 |H|^2\, \mathrm{d}\mu\le \frac{1}{m(m+1)}\int_{-1}^1 G\, \mathrm{d}\mu. \label{3.12}
\end{equation}
We refer to (\ref{3.12}) as the \emph{integral bound} for $|H|^2$.

Now we apply our two bounds (\ref{3.1}) and (\ref{3.12}) to (\ref{10}) and (\ref{13}). 
For the pointwise bound, substituting (\ref{3.1}) into (\ref{10}) gives
\begin{align}
(c_\mathrm{r}^2+c_\mathrm{i}^2)\int_{-1}^1G\, \mathrm{d}\mu&\le \int_{-1}^1\left\{\Omega^2-\beta^2+\frac{1-\mu^2}{m^2}\, 2\bigl[\beta(\mu\beta)'-\Omega(\mu\Omega)'\bigr]^+\right\}G\, \mathrm{d}\mu\,, \nonumber \\
&\le \left\{\Omega^2-\beta^2+\frac{1-\mu^2}{m^2}\, 2\bigl[\beta(\mu\beta)'-\Omega(\mu\Omega)'\bigr]^+\right\}_\mathrm{max}\int_{-1}^1G\, \mathrm{d}\mu\,,
\end{align}
where the plus sign superscript is defined by
\begin{equation}
f^+=\max(f,0).
\end{equation}
We have therefore arrived at the semicircle rule
\begin{equation}
c_\mathrm{r}^2+c_\mathrm{i}^2\le  \left\{\Omega^2-\beta^2+\frac{1-\mu^2}{m^2}\, 2\bigl[\beta(\mu\beta)'-\Omega(\mu\Omega)'\bigr]^+\right\}_\mathrm{max}. \label{3.5}
\end{equation}
Similarly, if we apply  (\ref{3.1}) to (\ref{13}) we derive another semicircle rule using the pointwise bound:
\begin{equation}
(c_\mathrm{r}-\overline{\Omega})^2+c_\mathrm{i}^2\le\left\{\Delta\Omega^2-\beta^2+\frac{1-\mu^2}{m^2}\, 2\bigl[\beta(\mu\beta)'-(\Omega-\overline{\Omega})(\mu\Omega)'\bigr]^+\right\}_\mathrm{max}. \label{3.8}
\end{equation}

To apply instead the integral bound, we first take the functions multiplying $G$ and $|H|^2$ out of the integral in equation (\ref{10}) and then use (\ref{3.12}). We obtain
\begin{align}
(c_\mathrm{r}^2+c_\mathrm{i}^2)\int_{-1}^1G\, \mathrm{d}\mu &\le (\Omega^2-\beta^2)_\mathrm{max}\int_{-1}^1G\, \mathrm{d}\mu+2\bigl[\beta(\mu\beta)'-\Omega(\mu\Omega)'\bigr]_\mathrm{max}\int_{-1}^1|H|^2\, \mathrm{d}\mu \nonumber\\
&\le (\Omega^2-\beta^2)_\mathrm{max}\int_{-1}^1G\, \mathrm{d}\mu+\frac{2\bigl[\beta(\mu\beta)'-\Omega(\mu\Omega)'\bigr]_\mathrm{max}^+}{m(m+1)}\int_{-1}^1G\, \mathrm{d}\mu.
\end{align}
Hence we have another semicircle rule
\begin{equation}
c_\mathrm{r}^2+c_\mathrm{i}^2\le (\Omega^2-\beta^2)_\mathrm{max}+\frac{2}{m(m+1)}  \bigl[\beta(\mu\beta)'-\Omega(\mu\Omega)'\bigr]_\mathrm{max}^+. \label{3.7}
\end{equation}
Similarly, applying the integral bound to equation (\ref{13}) yields
 \begin{equation}
(c_\mathrm{r}-\overline{\Omega})^2+c_\mathrm{i}^2\le (\Delta\Omega^2-\beta^2)_\mathrm{max}+\frac{2}{m(m+1)} \bigl[\beta(\mu\beta)'-\left(\Omega-\overline{\Omega}\bigr)(\mu \Omega)'\right]_\mathrm{max}^+. \label{3.9}
\end{equation}

In summary, we have derived four semicircle rules: (\ref{3.5}),  (\ref{3.8}), (\ref{3.7}) and (\ref{3.9}). \cites{Watson81} semicircle rule corresponds to (\ref{3.8}) with the field switched off (though the latter is tighter due to a more careful treatment of the geometric term).  
In the limit of small scale, $m\rightarrow \infty$, the geometric terms proportional to $(1-\mu^2)/m^2$ or ${1}/{m(m+1)}$  vanish and we recover the semicircle rules derived by \citet{Gilman67}, \citet{Cally00} and \citet{Hughes01} for MHD instability in Cartesian geometry, which can be further reduced to the theory of \citet{Howard62}    when the magnetic field is uniform.
When both limits are applied, (\ref{3.8}) and (\ref{3.9}) both reduce to the semicircle rule of \citet{Howard61}.  In that problem, the semicircle centred at $(c_\mathrm{r},c_\mathrm{i})=(0,0)$ completely includes the other  centred at $(\overline{\Omega},0)$ and  becomes redundant.

A distinguishing feature of the current theory is that for each possible semicircle centre, i.e.\ $(c_\mathrm{r}, c_\mathrm{i})=(0,0)$ and $(\overline{\Omega},0)$, there are two possible radii resulting from the two different methods used to bound $|H|^2$. The smaller radius will represent a tighter bound and thus be the effective one. We have found that, depending on the profiles of $\Omega$ and $\beta$, either of the two bounding methods can be tighter.
In general the results of the pointwise bounding, (\ref{3.5}) and (\ref{3.8}), are tighter when the shear of $\Omega$ or $\beta$ are prominent, because these bounds take the maximum of the sum of functions, in contrast to (\ref{3.7}) and (\ref{3.9}) which take the sum of the maxima of two functions. On the other hand, when the shears of $\Omega$ and $\beta$ are weak, the integral bounds (\ref{3.7}) and (\ref{3.9}) can be tighter due to the smaller coefficient of ${1}/{m(m+1)}$.

In an effort to make the results more compact and uniform, we present an alternative way to write the semicircle rules using functional expressions:
\refstepcounter{equation}
$$
c_\mathrm{r}^2+c_\mathrm{i}^2\le E[ f_1,g_1],\quad (c_\mathrm{r}-\overline{\Omega})^2+c_\mathrm{i}^2\le E[ f_2,g_2], \eqno(\theequation a,b)\label{3.20}
$$
where $E$ is a functional defined by
\begin{equation}
E\left[f,g\right]=\mathrm{min}\left(C[f,g],D[f,g]\right),
\end{equation}
with
\refstepcounter{equation}
$$
C[f,g]=\left\{f+\frac{1-\mu^2}{m^2}\, g^+\right\}_\mathrm{max},\quad D[f,g]=f_\mathrm{max}+\frac{1}{m(m+1)}\, g_\mathrm{max}^+,\eqno{(\theequation a,b)} \label{3.22}
$$
corresponding to the pointwise or integral bounds, respectively. The  functions in (\ref{3.20}) are
\begin{equation}
f_1=\Omega^2-\beta^2,\quad g_1=2\bigl[\beta(\mu\beta)'-\Omega(\mu\Omega)'\bigr],
\end{equation}
\begin{equation}
f_2=\Delta \Omega^2-\beta^2,\quad g_2=2\bigl[\beta(\mu\beta)'-(\Omega-\overline{\Omega})(\mu\Omega)'\bigr]. \label{3.24}
\end{equation}

We now demonstrate the application of  these semicircle rules to instability problems. Following \citet{Hughes01}, we first study the instability criterion. Since all the rules need to be satisfied by the complex value of $c$ for an arbitrary unstable mode, if the radius of either semicircle disappears, or if the semicircles for the two centres become disjoint, then instability is impossible. Using the notation in (\ref{3.20}), these conditions are
\refstepcounter{equation}
$$\label{3.25}
  E[f_1,g_1] \le0 \quad \mathrm{or}\quad  E[f_2,g_2] \le0 \quad  \mathrm{or} \quad  \sqrt{E[f_1,g_1]}+\sqrt{E[f_2,g_2]}\le|\overline{\Omega}|,
  \eqno{(\theequation a,b,c)}
$$
any of which is a sufficient condition for stability. A straightforward example of an application of (\ref{3.25}$b$) is the case of constant $\Omega$ and $\beta$:  the semicircle rule (\ref{3.9}) for this flow becomes
 \begin{equation}
(c_\mathrm{r}-\overline{\Omega})^2+c_\mathrm{i}^2\le \left[\frac{2}{m(m+1)}-1\right]\beta^2\le 0, \label{3.26}
\end{equation}
given that $m\ge 1$. Hence the flow is linearly stable when $\Omega$ and $\beta$ are constants. In fact, the phase velocities for this flow can be solved analytically, since the waves are spherical harmonics \citep{Marquez17}, but through this example we have shown the power of the semicircle rule: for this specific flow, it can give a bound that is tight enough to exclude the possibility of instability. We note that this analysis is based on equation (\ref{3.9}) which is found using the integral bound (itself tight for spherical harmonics); use of the pointwise bound is not tight enough to rule out instability for this flow.

\begin{figure}
  \centering
  \includegraphics[width=0.495\linewidth]{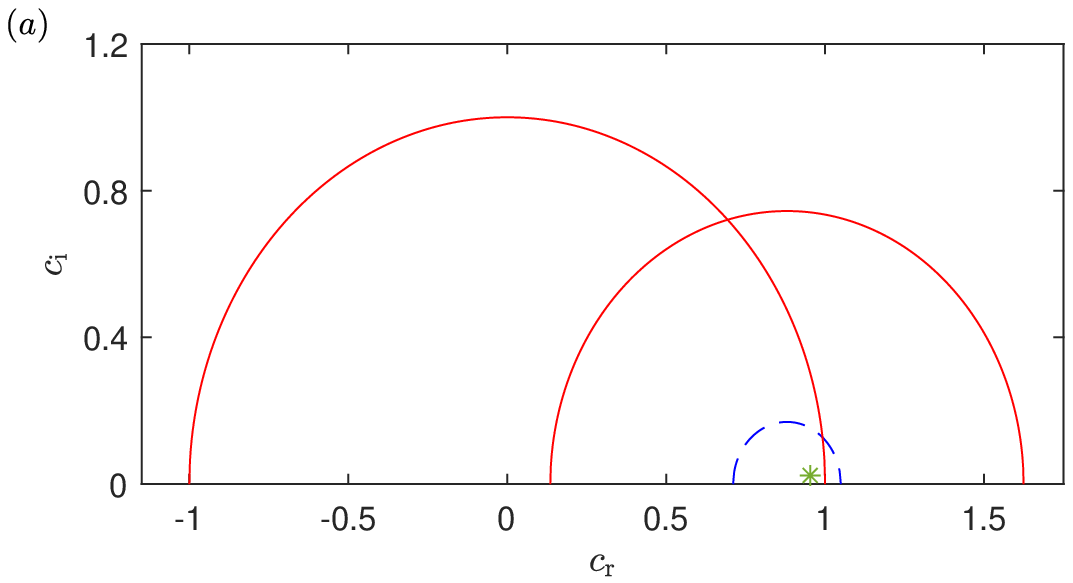}
  \includegraphics[width=0.495\linewidth]{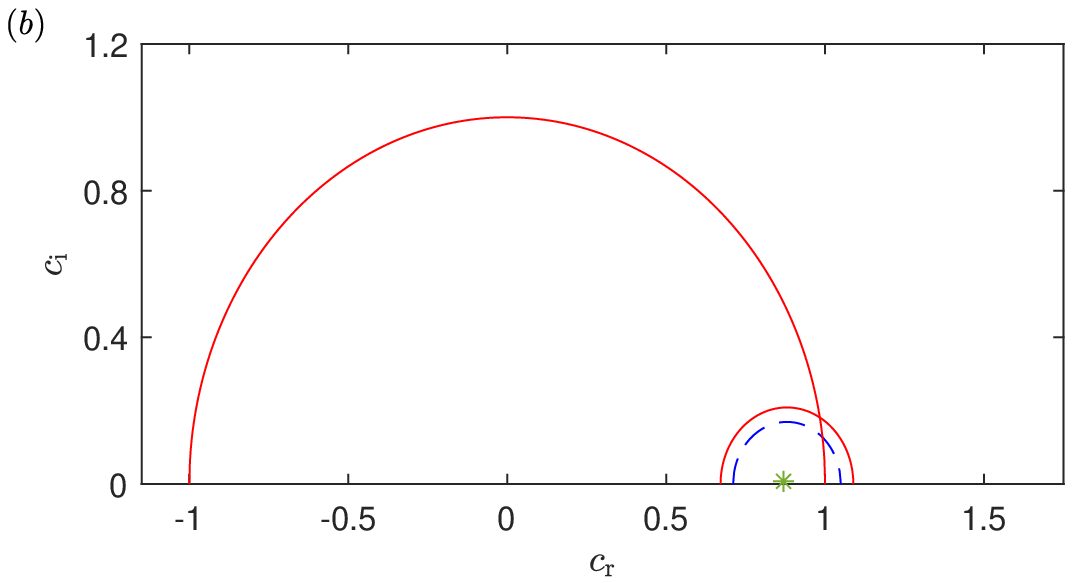}
  \caption{Semicircle rules for the zonal flow (\ref{0.1})  with $r=1$, $s=0.24$ and the magnetic field of ($a$) profile (\ref{3.28}) with $\sigma=1$ and ($b$) profile (\ref{3.29}) with  $\sigma=0.1$, $d=1/\sqrt{2}$ and $w = \pi/6$. The wavenumber is $m=1$ for both figures.  Solid lines represent the result of (\ref{3.5}) and (\ref{3.8}), and dashed lines represent the result of (\ref{3.8}) without the magnetic field. All semicircles plotted are the tightest, here from the pointwise bound.  The star represents the numerical solution for the unstable mode: $c=0.95+0.023\mathrm{i}$  for panel ($a$)  and $c=0.87+0.0074\mathrm{i}$ for panel $(b)$.
  }\label{F1}
\end{figure}

We recall that for the MHD problem in Cartesian geometry, for example, \cite{Hughes01}, the magnetic field may only reduce the radii of the semicircles, or keep them the same. Hence they explored how the field profile may realise one of (\ref{3.25}) to guarantee stability, and studied the tightness of these bounds.  In our spherical problem, on the other hand, the important feature is that the magnetic field may increase the radii of the semicircles through the geometric terms (those proportional to $(1-\mu^2)/m^2$ or ${1}/{m(m+1)}$),  and this is often accompanied by the destabilising effect of the magnetic field.  It is therefore of interest to show examples of the semicircles compared with the actual eigenvalues for sample profiles.  Motivated by the solar differential rotation profile, for the basic state angular velocity we consider the typical differential shear given by (\ref{0.1}). The parameters are chosen as $r=1$ and $s=0.24$ as in \citet{Gilman97}.
For the basic state magnetic field, we consider two examples: the first is a `linear shear' profile considered by \citet{Gilman97} and \citet{Cally01},
\begin{equation}\label{3.28}
 \beta=\sigma\mu,
\end{equation}
where $\sigma$ is a constant. 
The second is a profile with a pair of opposite Gaussian distributions,  studied by \citet{Dikpati99} and \citet{Cally01},
\begin{equation}
 \beta=\frac{\sigma}{2\sqrt{1-d^2}}\left[\exp\left( -\frac{4 (\mu-d)^2}{w^2(1-d^2)}\right)-\exp\left( -\frac{4 (\mu+d)^2}{w^2(1-d^2)}\right)\right], \label{3.29}
\end{equation}
where $\pm d$ are the centres of two Gaussian distributions and $w$ is a width parameter. These profiles are idealised models for the solar magnetic field. From solar magnetogram observations, the magnetic field is antisymmetric about the equator, and (\ref{3.28}) is the simplest antisymmetric profile while (\ref{3.29}) further models the belt patterns seen from observations. We plot the semicircles for the field profiles (\ref{3.28}) and (\ref{3.29}) in figures \ref{F1}$(a)$ and $(b)$ in solid lines, respectively. Mode $m=1$ is chosen, which is the only wavenumber where the flow is unstable. We choose $\sigma=1$ for  (\ref{3.28}) corresponding to a strong field, and $\sigma=0.1$ for (\ref{3.29}) corresponding to a weak field, and take $w = \pi/6$ and $d=1/\sqrt{2}$ for (\ref{3.29}).  Due to the strong shear in $\beta$, the semicircles (\ref{3.5}) and (\ref{3.8}) from use of the pointwise bound (\ref{3.22}$a$) have smaller radii, and so are plotted.
  We also plot the semicircle (\ref{3.8}) in the absence of a magnetic field in dashed lines; the other semicircle (\ref{3.5}) is the same with or without the magnetic field.  There is an unstable mode for each field profile, which is solved numerically and we plot as a star in each panel. As mentioned earlier, in the absence of magnetic field, the flow (\ref{0.1}) is unstable when $s/r>0.29$ \citep{Watson81}, hence the flow with $r=1$, $s=0.24$ that  is considered here is hydrodynamically stable, and the instability is induced by the magnetic field.

  The region where the two semicircles with solid lines overlap is the domain for all possible values of $c$.
   The star lies inside this region, as we expect.  For the strong field case of figure \ref{F1}$(a)$, the magnetic field significantly enlarges the original, hydrodynamic semicircle  (shown  dashed) centred at $\Omega=\overline{\Omega}$, though the actual unstable mode still lies inside this semicircle. For the weak field case of figure \ref{F1}$(b)$, the magnetic field slightly reduces the purely hydrodynamic semicircle, which is a little surprising since the field has a destabilising effect (the $\beta=0$ system is stable for this flow). 
In general, the situation of figure \ref{F1}$(a)$ (in which the magnetic field enlarges the semicircle centred at $\Omega=\overline{\Omega}$) is typical for field-induced instabilities. The opposite case (in which the field destabilises the flow but reduces the radius, as in figure \ref{F1}$(b)$) is relatively rare: it only happens for a weak field with certain profiles. As suggested by \citet{Gilman97}, the maximum possible growth rate from the semicircle rule can be much larger than the actual growth rate, but the rules are still powerful in giving rigorous bounds on $c$ in the complex plane. Note that in figure \ref{F1}$(a)$ the star is very close to the edge of the semicircle centred at $\Omega=0$, suggesting a tight bound in this case.

Further remarks on the situation seen in figure \ref{F1}$(a)$ may be helpful here. We note that the magnetic field increases one of the semicircles, but the eigenvalue still lies inside the original hydrodynamic semicircle.  The explanation is that the bounding of $H$ in terms of $|G|^2$ is over all possible functions, not necessarily the solutions of the eigenvalue problem; thus the bound can give a much larger domain than that attained by an actual eigenvalue.  We note that we have not found unstable modes outside the hydrodynamic semicircle,  and so we cannot answer the question of whether the larger semicircle induced by the magnetic field really represents a larger possible domain, or just a looser bound.  We leave this issue for future consideration.


We give another example of the prediction of the semicircle rules for profiles that are related to the analysis of the clamshell instability in the subsequent sections.
We consider a  magnetic field with $\beta=0$ at a certain latitude, which coincides with the location where $|\Omega|$ reaches its maximum. The standard profiles (\ref{0.1}) with (\ref{3.28}) or (\ref{3.29}) studied above have this property when $rs>0$ (the case for solar differential rotation). It can be shown from (\ref{3.20})--(\ref{3.24}) that
        \begin{equation}\label{3.30}
      E[f_1, g_1] \ge ({\Omega^2})_\mathrm{max},\quad E[f_2,g_2] \ge \Delta\Omega^2.
     \end{equation}






If $\Omega$ is not a constant function of latitude, in other words the fluid flow is sheared, then using (\ref{3.30}) it is evident that the sufficient conditions for stability given in (\ref{3.25}) are never satisfied. In fact, the lower limits given by (\ref{3.30}) are the results of \cite{Howard61}, where there is a finite-area semicircle for any sheared flow in the plane. A key observation here is that the semicircle rules we have obtained always allow the possibility of instability in a sheared flow on the sphere provided the magnetic field vanishes at the latitude where the rotation rate is greatest. We will demonstrate that such an instability does indeed exist,  in the limit of weak shear, by our analysis in \S \ref{S5.1}.

We finally note that in addition to the theory we present, there could be other versions of semicircle rules. Our derivation is based on the equation for $H=\psi/(\Omega-c)$ in Sturm–Liouville form, following the approach of \citet{Watson81}, but \citet{Thuburn96} and \citet{Sasaki12} worked on the Sturm–Liouville equation of  $\eta=\psi/[(\Omega-c)\sqrt{1-\mu^2}]$ and derived semicircle rules different from Watson's. At present, we do not know which version is tighter. Indeed, there could more versions that result from other substitutions. For MHD instability in Cartesian geometry, \citet{Deguchi21} improved the traditional semicircle rules by finding the inner envelope of a family of semicircles, and so a yet tighter bound. These approaches could provide avenues to extend our theory, and we leave them for future study.

\section{Analysis of the clamshell instability} \label{asymptotics}

\begin{figure}
  \centering
   \includegraphics[width=0.6\linewidth]{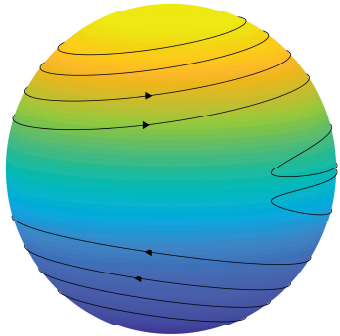}
  \caption{Magnetic field lines for the instability of the basic state $\Omega=1-0.1\mu^2$, $\beta=\mu$ with wavenumber $m=1$.  The eigenvalue is obtained numerically as $c=0.982+0.00847\mathrm{i}$.  The plotted results correspond to the basic state plus the unstable mode with a small amplitude  shown in figure \ref{F2}.  
  }\label{F0}
\end{figure}

\citet{Cally01} and \citet{Cally03} have shown that when the magnetic field is strong and its profile is broad,  MHD instability on a sphere  features a clamshell pattern in which the field lines on the two hemispheres are tilted in opposite directions. In figure \ref{F0} we give an example of such a clamshell pattern, which corresponds to the basic state (\ref{0.1}) and (\ref{3.28}) plus the unstable mode with its eigenfunction shown in figure \ref{F2}. Without loss of generality, we normalise the eigenfunction by $\max |H|=0.07$ and $\ImIm H=0$ as $\mu\rightarrow 1$ throughout the paper; other normalisations would yield a similar pattern to figure \ref{F0} as long as $|H|$ remains small.

In this section we will undertake an asymptotic analysis of this instability in the limit of weak shear $\Omega'$ of the basic zonal flow $\Omega$.
We will derive an asymptotic solution for the eigenvalue $c$ in this limit, which will provide insights into the instability mechanism applying to a wide range of profiles of $\Omega$ and $\beta$.  

\subsection{The tilting mode}

While equations (\ref{1.27}),  (\ref{2.24}) and (\ref{2.27}) look quite complicated, for  azimuthal wavenumber $m=1$ they admit a very simple solution for arbitrary $\Omega$ and $\beta$, namely
\begin{equation}
c=0,\quad \psi=\sqrt{1-\mu^2}\, \Omega,\quad \chi=\sqrt{1-\mu^2}\, \beta,\quad H=\sqrt{1-\mu^2}\, . \label{4.1}
\end{equation}
The physical meaning of this solution lies in the rotational invariance of the spherical geometry: if we apply a solid-body rotation to the entire zonal flow and magnetic field, the tilted flow and field  are still solutions of the governing equations. When the angle of tilting is small, the solution given in (\ref{4.1}) gives the difference between the tilted and original states. 
To see this, first note that the rotation through an infinitesimal angle $\alpha$ about the $x$-axis is given by $(x,y,z) \to(x,y - \alpha z, z+\alpha y)$ and corresponds to the spherical polar coordinate change
\begin{equation}
\theta \to \theta - \alpha \sin \phi, \quad \phi \to \phi - \alpha \cot \theta \cos \phi.
\end{equation}
Thus any function $f(\theta, \phi)$ is mapped under such a rotation by
\begin{equation}
f(\theta, \phi) \to f(\theta + \alpha \sin \phi, \phi + \alpha \cot \theta \cos \phi) \simeq f (\theta,\phi) + \alpha [ \sin \phi \, f_\theta + \cot\theta \cos \phi \, f_\phi] .
\end{equation}
Apply this to a stream function $\Psi(\theta)$ for the basic state, satisfying $U = \Omega\, \sin \theta = - \Psi_\theta$   (cf. (\ref{2.20})), and we find that the tilted stream function is
\begin{equation}
\Psi \to \Psi + \alpha \sin \phi \, \,\Psi_\theta = \Psi - \alpha \sin \phi \,  \sin \theta\, \Omega =  \Psi  + \left(\tfrac{1}{2} \mathrm{i} \alpha  \sqrt{1-\mu^2}\, \Omega \, e^{\mathrm{i} \phi} + \mathrm{c.c.}\right).
\end{equation}
The difference corresponds to a multiple $\tfrac{1}{2}\mathrm{i}\alpha$ of the steady solution (\ref{4.1}) to the linear problem (and likewise a multiple $\tfrac{1}{2}\alpha$ is a rotation around the $y$-axis).
For hydrodynamic stability, this neutral mode was noticed by  \citet{Watson81} (although there appears to be a mistake in the form of the eigenfunction given). A well-known analogue is the neutral mode arising from the translational invariance of a
vortex in the plane: shifting the entire $m=0$ vortex gives an $m=1$ solution to the linear problem with zero eigenvalue, as noted by \citet{Bernoff94}.


Interestingly, for the MHD problem, equations (\ref{1.27}),  (\ref{2.24}) and (\ref{2.27}) also admit a slightly different mode that involves the tilting of the basic state. If $\Omega$ is a constant, i.e. the zonal flow is solid body rotation, then for arbitrary $\beta$ we also have an exact solution for $m=1$:
\begin{equation}
  c=\Omega, \quad \psi=0, \quad \chi=\sqrt{1-\mu^2}\, \beta, \quad H=\sqrt{1-\mu^2}\, . \label{4.2}
\end{equation}
In this solution there is no velocity disturbance and the magnetic field lines are slightly tilted, similar to (\ref{4.1}), but now they rotate with the solid body rotation $\Omega(\mu)=c$, frozen into the flow. Furthermore, if the magnetic field vanishes somewhere, say
  \begin{equation}\label{4.3}
    \beta=0\quad {\mathrm{at}} \quad \mu=\mu_\star,
  \end{equation}
then equation (\ref{2.30a}) is satisfied and $\mu_\star$ is a critical level where equation (\ref{2.27}) becomes singular. However, the solution (\ref{4.2}) remains regular there. The mathematical picture is that (\ref{2.27}) has two linearly independent solutions: one is singular at $\mu=\mu_\star$,  the other is regular there and is the solution (\ref{4.2}). %

\begin{figure}
  \centering
  \includegraphics[width=0.495\linewidth]{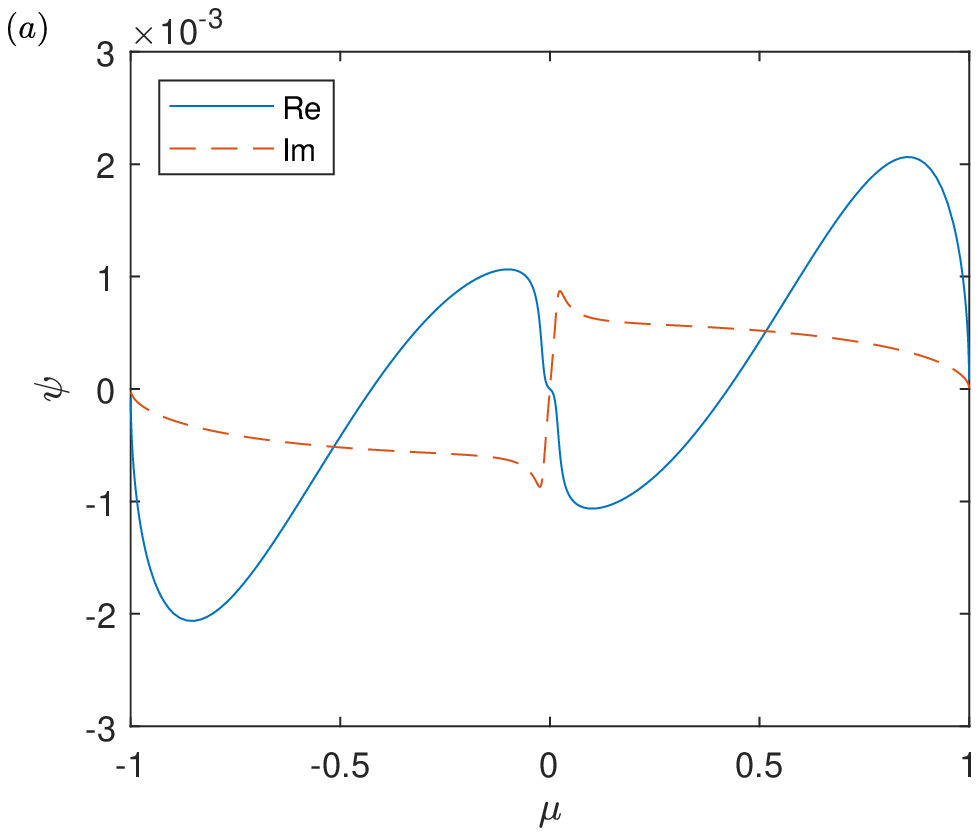}
   \includegraphics[width=0.495\linewidth]{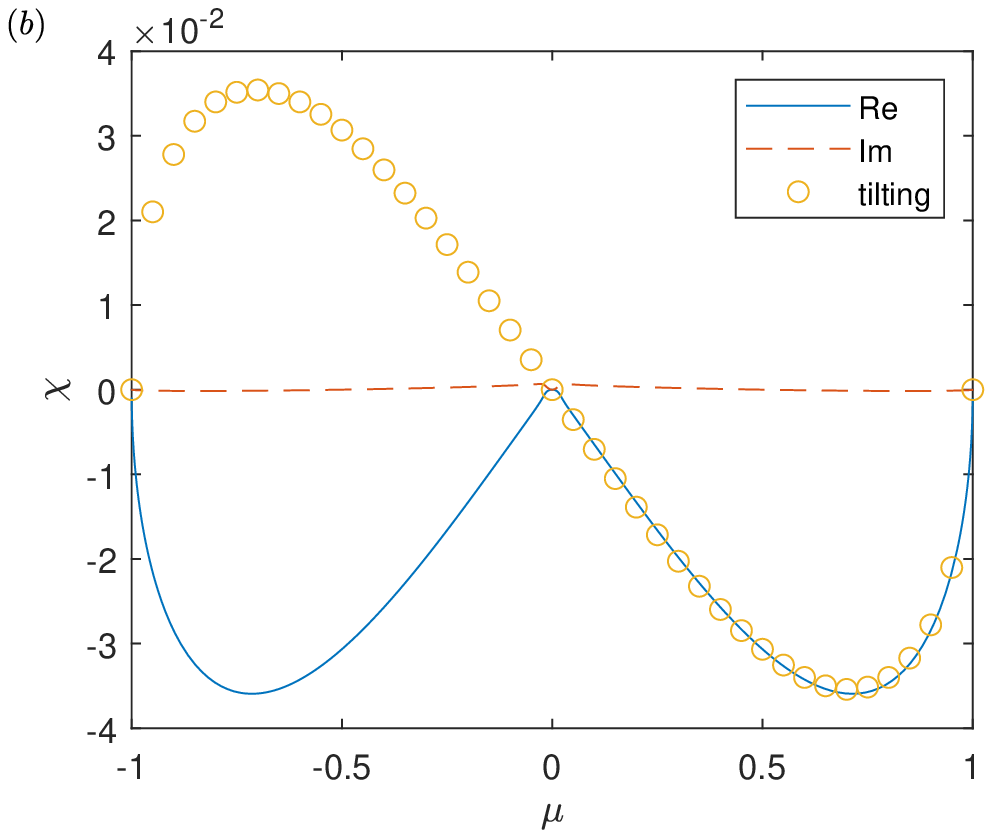}
      \includegraphics[width=0.495\linewidth]{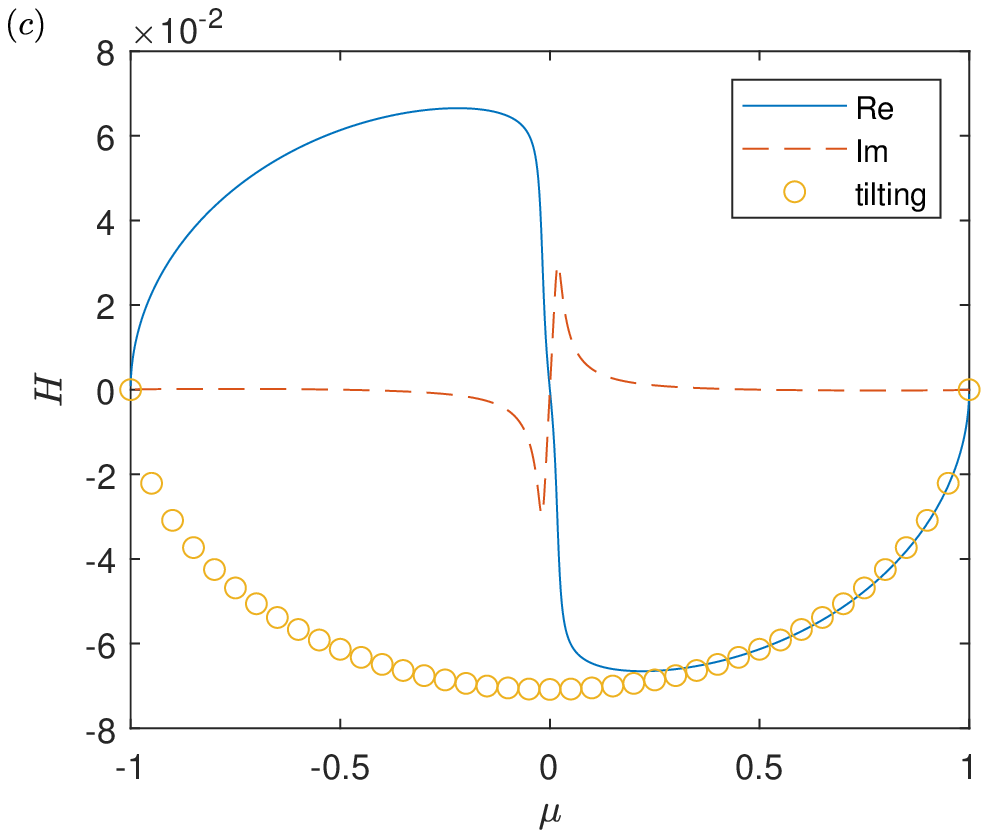}
  \caption{Eigenfunctions for the instability of the profile given in figure \ref{F0}. The tilting mode (\ref{4.2}) is plotted using circles. The eigenfunctions are normalised by $\max |H|=0.07$ and $\ImIm H=0$ as $\mu\rightarrow 1$.  }\label{F2}
\end{figure}

The solution (\ref{4.2}) is closely related to the clamshell instability.  From now on, we refer to it as the \emph{tilting mode}.  For the example shown in figures \ref{F0} and \ref{F2}, the shear of $\Omega$ is weak, and so we expect that the tilting mode (\ref{4.2}), derived for constant $\Omega$, is relevant.  In figure \ref{F2}, we plot the eigenfunctions that correspond to the clamshell pattern in figure \ref{F0} together with (\ref{4.2}).  
 We see that in most of the region the tilting mode agrees very well with the unstable mode; however, interestingly, the latter is reversed across $\mu=0$. It turns out that the weak shear in $\Omega$ excites the singular solution, which becomes dominant at the critical level; this is at $\mu_\star=0$ for the magnetic profile $\beta=\mu$. The singularity has a dramatic impact on the eigenfunction: it reverses the sign of the tilting mode, opening up the clamshell in figure \ref{F0}, and importantly it destabilises the flow. Based on this intuition, we will undertake an asymptotic analysis to solve the eigenvalue problem in  \S 4.2.  Readers who are not interested in the technical details of the matched asymptotic expansions may jump to the final result given by equation (\ref{4.21}), noting that a star subscript represents the value of a function evaluated at $\mu=\mu_\star$.

It is noted that we have not yet found a normal mode instability that is induced by  the mode (\ref{4.1}).  This is perhaps because this mode is `too stable': the eigenvalue $c=0$ remains unchanged however $\Omega$ and $\beta$  vary. Note that when $\Omega=0$, (\ref{4.1}) and (\ref{4.2}) become identical, and then our analysis indicates that the resonance between the two modes may induce algebraic growth (instead of exponential growth) of the magnetic field. But given $\Omega=0$ is not very common in astrophysics, this is of limited interest,  and we will not discuss it further.


\subsection{The matched asymptotic expansions}

We consider profiles of $\Omega(\mu)$ and $\beta(\mu)$ such that the variation in $\Omega$ is weak and $\beta$ is zero at a location denoted by $\mu=\mu_\star$. For simplicity, in the present study we assume that there is only one such $\mu_\star$ in the domain (although the case of multiple $\mu_\star$ may be studied following the same method). We also take $\beta$ to pass through zero at $\mu=\mu_\star$ with a gradient that is not small. Thus, close to $\mu_\star$, we have the approximation
\begin{equation}\label{4.4a}
  \beta\approx \beta_\star'(\mu-\mu_\star),\quad \mu\approx \mu_\star,
\end{equation}
where $\beta_\star'$ is $\beta'$ evaluated at $\mu=\mu_\star$, and we assume that $\beta_\star'$  is of order unity (or larger). Apart from these prescriptions,  $\Omega(\mu)$ and $\beta(\mu)$ are general functions.
 We perform matched asymptotic expansions that combine the bulk of the flow, where the disturbance is mainly represented by the tilting mode, and the critical layer near $\mu=\mu_\star$, where the solution changes rapidly.  The azimuthal wavenumber will be fixed at $m=1$,  since this is the only wavenumber that admits the tilting mode.

 We express $\Omega$ and $c$ by
 \begin{equation}\label{4.4}
  \Omega=\Omega_0+\varepsilon \Omega_1 (\mu),\quad c=\Omega_0+\varepsilon c_1  + \cdots.
 \end{equation}
Here the leading order rotation is $\Omega_0\neq0 $, which is a constant representing the angular velocity of the solid body rotation, while $\varepsilon$ is a small number representing the amplitude of the weak shear, and  $\Omega_1$ is an arbitrary function representing the shear profile.  The eigenvalue $c$ is $\Omega_0$ at leading order, following the tilting mode (\ref{4.2}). The weak shear $\varepsilon \Omega_1$ induces a small correction $\varepsilon c_1$ to the eigenvalue, and our goal is to determine this.
Also, if $\Omega_0=0$ in (\ref{4.4}), but $\Omega$ as a whole is weak compared to $\beta$, we can still follow a similar asymptotic analysis. This situation is perhaps less relevant for astrophysical applications and so the derivation is consigned to appendix A; it is different in detail, but the final result  (\ref{4.21}) is the same.



\subsubsection{Outer solution} \label{4.2.1}



Away from the critical level $\mu=\mu_\star$, we expand $H$ as
\begin{equation}\label{4.5}
  H=H_0(\mu)+\varepsilon H_1(\mu) +\cdots\,,
\end{equation}
where $H_0$ has the profile of the tilting mode but is discontinuous across the critical layer,  as we see in figure \ref{F2}. Thus,
\begin{equation}
H_0(\mu)=\left\{\begin{array}{ll}
A_-\sqrt{1-\mu^2}\, , \quad \mu<\mu_\star, \\
[5pt]
A_+\sqrt{1-\mu^2}\, ,\quad \mu>\mu_\star\,,
\end{array}\right. \label{4.6}
\end{equation}
where $A_+$ and $A_-$ are constants. $H_1$ represents the small correction that is induced by the weak shear in $\Omega$. Substituting (\ref{4.4}) and (\ref{4.5}) into (\ref{2.27}) and collecting terms at $O(\varepsilon)$, we obtain the equation for $H_1$,
\begin{equation}
\left[\beta^2(1-\mu^2)H_1'\right]'+\left[2\beta(\mu\beta)'-\frac{\beta^2}{1-\mu^2}\right]H_1 =2\Omega_0(\Omega_1-c_1)H_0. \label{4.7}
\end{equation}

Using the method of variation of constants, noting that $H_1$ must be finite at $\mu=\pm1$,  the solution for $H_1$ can be obtained in the form
\begin{equation}\label{4.7a}
  H_1(\mu)=\left\{\begin{array}{lll} h_{\mathrm{s}}(\mu)\displaystyle \int_{ 1}^\mu2\Omega_0\left[\Omega_1(\nu)-c_1\right]H_0(\nu)h_{\mathrm{r}}(\nu)\, \mathrm{d}\nu  \\
  \qquad -h_{\mathrm{r}}(\mu)\displaystyle\int_1^\mu 2\Omega_0[\Omega_1(\nu)-c_1]H_0(\nu)h_{\mathrm{s}}(\nu)\, \mathrm{d}\nu+C_a h_\mathrm{r}(\mu), \quad \mu_\star<\mu<1,   \\
  [10pt]
  h_{\mathrm{s}}(\mu)\displaystyle\int_{- 1}^\mu2\Omega_0\left[\Omega_1(\nu)-c_1\right]H_0(\nu)h_{\mathrm{r}}(\nu)\, \mathrm{d}\nu \\
  \qquad -h_{\mathrm{r}}(\mu)\displaystyle\int_{-1}^\mu 2\Omega_0[\Omega_1(\nu)-c_1]H_0(\nu)h_{\mathrm{s}}(\nu)\, \mathrm{d}\nu+C_bh_\mathrm{r}(\mu), \quad -1<\mu<\mu_\star,
  \end{array}\right.
\end{equation}
where  $h_{\mathrm{r}}$ and $h_{\mathrm{s}}$ are the regular and singular solutions of the corresponding homogeneous equation, given by
\begin{eqnarray}
&&h_{\mathrm{r}}(\mu)=\sqrt{1-\mu^2}\, ,\quad h_{\mathrm{s}}(\mu)=\left\{\begin{array}{ll}
                                                                    \sqrt{1-\mu^2}\displaystyle\int_{\mu_a}^\mu\frac{1}{\beta(\nu)^2(1-\nu^2)^2}\, \mathrm{d}\nu\qquad \mu_\star<\mu,\mu_a<1, \\
                                                                      [10pt]
                                                                    \sqrt{1-\mu^2}\displaystyle \int_{\mu_b}^\mu\frac{1}{\beta(\nu)^2(1-\nu^2)^2}\, \mathrm{d}\nu\quad \;  -1<\mu,\mu_b<\mu_\star\,,
                                                                  \end{array} \right.   \nonumber \\ \label{4.8}
\end{eqnarray}
where $\mu_a$ and $\mu_b$ may be arbitrarily chosen in the given range,  and  $C_a$ and $C_b$ are undetermined constants, a consequence of $h_\mathrm{r}$ satisfying both boundary conditions.
%
%
Using (\ref{4.4a}), we can deduce that $H_1, h_{\mathrm{s}}\sim(\mu-\mu_\star)^{-1}$ as $\mu\rightarrow \mu_\star$. Hence  unlike the leading order term $H_0$, the correction $H_1$ is essentially singular at $\mu=\mu_\star$. In other words, combining the tilting mode $H_0$ (in either hemisphere) with weak shear excites the singular mode of the system.  When $\mu-\mu_\star=O(\varepsilon)$, $\varepsilon H_1$ becomes as large as $H_0$ and the expansion (\ref{4.5}) becomes disordered. We note the singular behaviour of $H_1'$:
\begin{equation}
\!\!\!H_1'=\frac{A_\pm}{(1-\mu_\star^2)^{{3}/{2}}\beta_{\star}'^2(\mu-\mu_\star)^2}\int_{\pm 1}^{\mu_\star} \!2\Omega_0\left[\Omega_1(\mu)-c_1\right](1-\mu^2)\, \mathrm{d}\mu+O\Bigl(\frac{1}{\mu-\mu_\star}\Bigr),\quad \mu\rightarrow \mu_\star^{\pm}, \label{4.9}
\end{equation}
which we will use later.

\subsubsection{Inner solution}
The region $\mu-\mu_\star =O(\varepsilon)$ is the critical layer where $H$ varies significantly.   We therefore introduce a local stretched coordinate, writing
\begin{equation}
 \eta=\frac{\mu-\mu_\star}{\varepsilon}\, ,\qquad H=\mathcal{H}(\eta) + \cdots. \label{4.10}
\end{equation}
Substituting (\ref{4.4a}), (\ref{4.4}) and  (\ref{4.10}) into (\ref{2.27}) we find that the leading order   local equation is
\begin{equation}\label{4.12}
   \frac{\mathrm{d}}{\mathrm{d}\eta}\biggl\{\left[(\Omega_{1\star}-c_1)^2-\beta_\star'^2\eta^2\right] \frac{\mathrm{d}\mathcal{H}}{\mathrm{d }\eta}\biggr\}=0,
\end{equation}
where $\Omega_{1\star}$ is $\Omega_1$ evaluated at $\mu=\mu_\star$. Integrating with respect to $\eta$, we obtain
\begin{equation}\label{4.13}
 \frac{\mathrm{d} \mathcal{H}}{\mathrm{d}\eta}=\frac{\alpha_1}{(\Omega_{1\star}-c_1)^2-\beta_\star'^2\eta^2 }\, ,
\end{equation}
and hence
\begin{equation}\label{4.14}
  \mathcal{H}=\frac{\alpha_1}{2(\Omega_{1\star}-c_1)|\beta_\star'|}\, \bigl[\log\left(|\beta_\star'|\eta+\Omega_{1\star}-c_1\right)-\log\left(|\beta'_\star|\eta-\Omega_{1\star}+c_1\right)\bigr]+\alpha_2,
\end{equation}
for some constants $\alpha_1$ and $\alpha_2$.

A striking property of the critical layer is the presence of a significant jump in $H$ from one side to the other, which is shown in figure \ref{F2}$(c)$ (also see the later figure \ref{F_comparison}). Such a jump can be understood from the asymptotic solution (\ref{4.14}) as follows. If we select the branch cuts of the logarithm functions to lie on the negative real axis, then for $\ImIm c_1>0$ their large-variable limits are
\refstepcounter{equation}
$$
  \log\bigl[|\beta_\star'|\eta\pm(\Omega_{1\star}-c_1)\bigr]\sim\left\{\begin{array}{ll}
                                                                              \log|\beta_\star'\eta|,\qquad \quad\, \eta\rightarrow +\infty, \\
                                                                              [5 pt]
                                                                                \log|\beta_\star'\eta|\mp \mathrm{i}\pi,\quad\eta\rightarrow -\infty.
                                                                              \end{array}\right. \eqno{(\theequation a,b)} \label{logJump}
$$
Hence
\refstepcounter{equation}
$$
\mathcal{H}\sim\left\{\begin{array}{ll}
\alpha_2, \qquad\qquad \qquad \qquad\qquad\;\; \eta\rightarrow+\infty, \\
[5pt]
\displaystyle \frac{-\mathrm{i}\pi\alpha_1}{(\Omega_{1\star}-c_1)|\beta_\star'|}+\alpha_2, \quad\qquad \eta\rightarrow-\infty.
\end{array}\right.     \eqno{(\theequation a,b)}\label{Hjump}
$$
The first term of (\ref{Hjump}$b$) clearly indicates the jump of $\mathcal{H}$ across the critical layer. This jump is important because it reverses the tilting direction to form the clamshell pattern, and induces instability through the presence of the imaginary unit $\mathrm{i}$, as we will see later. According to (\ref{logJump}),  the jump of the logarithm functions is contingent on the existence of a non-zero $|\beta_\star'|$, highlighting the role of the gradient of the magnetic field at the critical level. 


\begin{figure}
  \centering
   \includegraphics[width=0.75\linewidth]{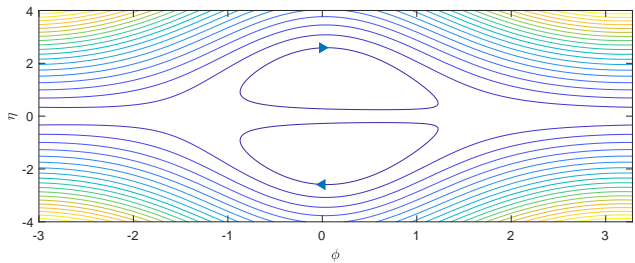}
  \caption{The magnetic field lines in the critical layer, corresponding to figure \ref{F0} near the equator. We have superimposed the basic magnetic field with the asymptotic local solution (\ref{4.14}), where $\alpha_1$ and $\alpha_2$ are found by matching to $A_-$ and $A_+$ via (\ref{4.16}) and $(\ref{4.19})$, and $A_-$ and $A_+$ are obtained by fitting to the numerical solution shown in figure \ref{F_comparison}.  }\label{F_inner}
\end{figure}

The magnetic field lines in the critical layer, rendered by the local solution (\ref{4.14}) plus the basic magnetic field, are shown in figure \ref{F_inner}. It may be seen that the critical layer induces a pair of closed loops in the field line pattern, which is also visible in figure \ref{F0} near the equator. Note that such a pattern was not shown in the corresponding figure  of  \cite{Cally01} (first and second panel of his figure 4), since he did not draw field lines in the critical layer in the early stage of the evolution. Also note that at later times,  \cites{Cally01} simulation has shown that the field lines on the two hemispheres will reconnect as a result of dissipation. This is quite different from the ideal MHD instability that we currently study: in our figure \ref{F_inner},  the field lines on the two hemispheres are separated.


\subsubsection{ Matching and eigenvalue} \label{4.2.3}

Matching the inner and outer solution provides relations between the constants $\alpha_1$, $\alpha_2$, $A_-$ and $A_+$, and so determines the eigenvalue $c_1$. We first match $H'$ from the inner and outer solution in an intermediate region $\mu-\mu_\star=O(\varepsilon^{{1}/{2}})$. Here $H_0'\ll \varepsilon H'$, so that we can neglect the former in the outer solution. Hence matching (\ref{4.13}) and (\ref{4.9}) via
\begin{equation}\label{4.15}
  \frac{1}{\varepsilon}\frac{\mathrm{d}\mathcal{H}}{\mathrm{d}\eta}\,  \bigg|_{\eta\rightarrow \pm \infty}= \varepsilon H_1' \, \bigg|_{\mu\rightarrow \mu_\star^\pm}\, ,
\end{equation}
we find
\refstepcounter{equation}
$$\label{4.16}
\alpha_1 =\frac{A_-}{(1-\mu_\star^2)^{{3}/{2}}}\int^{-1}_{\mu_\star}2\Omega_0[\Omega_1(\mu)-c_1](1-\mu^2)\, \mathrm{d}\mu  \nonumber \\
$$
$$
\qquad =\frac{A_+}{(1-\mu_\star^2)^{{3}/{2}}}\int^{1}_{\mu_\star}2\Omega_0[\Omega_1(\mu)-c_1](1-\mu^2)\, \mathrm{d}\mu\,, \eqno{(\theequation a,b)}
$$
this providing two relations between $\alpha_1$, $A_+$ and $A_-$. Next, we match $H$: at $\mu-\mu_\star=O(\varepsilon^{{1}/{2}})$,  the outer solution of $H$ is dominated by the tilting mode $H_0$, so that the matching condition is
\begin{equation}\label{4.17}
  \mathcal{H} \, \big|_{\eta\rightarrow \pm \infty}=H_0 \, \big|_{\mu\rightarrow \mu_\star^\pm}\, .
\end{equation}
According to (\ref{4.6}) and (\ref{Hjump}),
\refstepcounter{equation}\label{4.19}
$$
 \alpha_2=A_+\sqrt{1-\mu_\star^2}\, ,
\qquad
\frac{-\mathrm{i}\pi\alpha_1}{(\Omega_{1\star}-c_1)|\beta_\star'|}+\alpha_2=A_-\sqrt{1-\mu_\star^2}\, . \eqno{(\theequation a,b)}
$$
\begin{figure}
  \centering
 \includegraphics[width=0.58\linewidth]{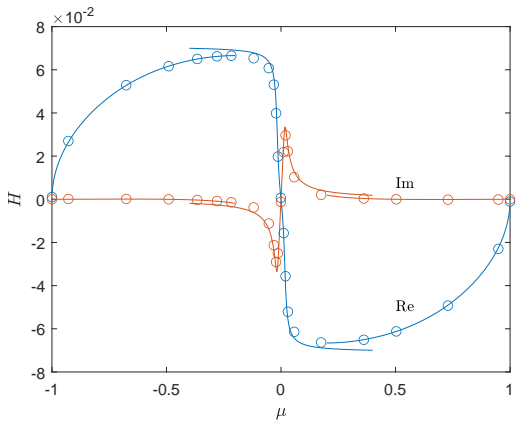}
  \caption{The comparison of the eigenfunction $H$ between  the asymptotic solution (solid lines) and numerical solution (circles) for the unstable mode of figure \ref{F2}.  The asymptotic solution of $H$ consists of the outer solution (\ref{4.5}), (\ref{4.6}), (\ref{4.7a}) and the inner solution (\ref{4.14}),  with the matching condition (\ref{4.16}) and (\ref{4.19}).  The eigenvalue $c$ of the asymptotic solution (\ref{4.21}) is $0.984+0.00778\mathrm{i}$,  while its numerical solution is $0.982+0.00847\mathrm{i}$.
   }\label{F_comparison}
\end{figure}

Combining (\ref{4.16}) and (\ref{4.19}), a non-trivial solution for $\alpha_1$, $\alpha_2$, $A_-$ and $A_+$  yields an equation which determines the eigenvalue $c_1$ for $\ImIm c_1>0$:
\begin{align}\label{4.20}
&  \frac{1}{\displaystyle\int_{-1}^{\mu_\star} 2\Omega_0 [c_1-\Omega_1(\mu)](1-\mu^2)\,\mathrm{d}\mu}+  \frac{1}{\displaystyle\int_{\mu_\star}^1 2\Omega_0 [c_1-\Omega_1(\mu)](1-\mu^2)\,\mathrm{d}\mu} \nonumber \\
& \qquad \qquad\qquad\qquad =\frac{\mathrm{i}\pi}{|\beta_\star'|(1-\mu_\star^2)^2(c_1-\Omega_{1\star})}\, .
\end{align}
An example of the comparison between the asymptotic and numerical solutions is shown in figure \ref{F_comparison}.   Note that if (\ref{4.20}) yields a solution with $\ImIm c_1<0$, it is not a valid normal mode solution, since it contradicts our branch cut selection (\ref{logJump}) based on $\ImIm c_1>0$.   Using $\ImIm c_1<0$ for the branch cut selection would result in a solution with $\ImIm c_1>0$, which is again a contradiction.    In this case,  the normal mode disappears due to the excitation of the critical layer.  In similar problems of hydrodynamic stability theory,  it is possible to consider an initial value problem to recover part of the behaviour of such a mode,  known as a `quasi-mode'  \citep{Briggs70},  but we will not consider this problem in this paper.





It is useful to express (\ref{4.20}) using the original variables $c$ and $\Omega$ instead of $c_1$ and $\Omega_1$ via (\ref{4.4}). The resulting equation is presented as (\ref{4.21}) at the beginning of the next section, where we also discuss its implications. This is an equation for the eigenvalue $c$ with $c_\mathrm{i}>0$, and with $\Omega_\star$ equal to $\Omega(\mu_*)$. We have replaced $\Omega_0=\Omega_\star$ for the leading-order solid body rotation; as the shear is weak, $\Omega$ is approximately constant everywhere. Equation (\ref{4.21}) is derived under the condition that $\Omega$ and $\beta$ are of the same order and  the shear of $\Omega$ is small compared to both of these. In appendix A, we also present the analysis for the situation where $\Omega$ as a whole is small compared to $\beta$. 
 The derivation is a little different but the final result remains the same as (\ref{4.21}), and so this equation is generally applicable as long as the shear $\Omega'$ is weak compared to the magnetic field $\beta$.


\section{Results and discussion} \label{discussion} \label{S5}



The result of our analysis is the following implicit equation for the complex wave speed $c$, taken to have a positive imaginary part $c_{\mathrm{i}}>0 $ that gives the growth rate of a mode:
\begin{equation}\label{4.21}
  \frac{c-\Omega_\star}{\displaystyle\int_{-1}^{\mu_\star} [c-\Omega(\mu)] (1-\mu^2)\,\mathrm{d}\mu}+  \frac{c-\Omega_\star}{\displaystyle\int_{\mu_\star}^1 [c-\Omega(\mu)](1-\mu^2)\,\mathrm{d}\mu}=\frac{2\mathrm{i}\pi\Omega_\star}{|\beta_\star'|(1-\mu_\star^2)^2}\, .
\end{equation}
Here  we recall that the magnetic field profile $\beta(\mu)$ has a single, simple zero at the critical latitude given by $\mu = \mu_*$, where the gradient $\beta'_* \neq 0$  and the angular velocity is $\Omega_*$. The equation is valid provided the shear $\Omega'(\mu)$ of the angular velocity profile is small compared with the magnetic field.

\subsection{ General results} \label{S5.1}
Equation (\ref{4.21}) has a relatively simple form and we can use it to gain significant insights into the instability properties for general profiles of $\Omega$ and $\beta$. First, we observe that $\beta$ only enters this equation through $\beta'_\star=\beta'(\mu_\star)$, where $\mu_\star$ is the location where $\beta=0$. The other properties of $\beta$ (e.g.~the value of $\beta$ at other latitudes) do not affect (\ref{4.21}). This is a curious property, because the MHD instability is global, yet the magnetic field only affects the instability through its local behaviour in the critical layer. To test this finding, we consider three different profiles: $\beta=\mu$, $\sin\mu$ and $e^\mu-1$. They all have $\mu_\star=0$ and $\beta'_\star=1$, and so the same asymptotic result for $c$ given by  (\ref{4.21}). For the zonal flow, we select $\Omega=1-s\mu^2$ with $s=0.12$ and $s=0.06$; the case of  $s=0.12$ has been used by \citet{Cally01}  as a model for solar differential rotation.
The results of the eigenvalues are displayed in table~\ref{T1}. For comparison, the asymptotic solution (\ref{4.21}) is given in the last row. We see that the eigenvalues $c$ for the various profiles are indeed close, and interestingly, to a much higher degree than the precision of the asymptotic solution.    We summarise this conclusion as follows: \textit{the magnetic field profile only affects the instability through the location where it passes though zero and the value of its gradient there. }


\begin{table}
  \begin{center}
\def~{\hphantom{0}}
  \begin{tabular}{lccc}
                               & $s=0.12$   &   $s=0.06$   \\[3pt]
       $\beta=\mu$   & $0.9780+ 1.037\times 10^{-2}\mathrm{i}$& ~~$0.9895+4.897\times 10^{-3}\mathrm{i}$~ \\
       $\beta=\sin \mu$   &$0.9780+1.035\times 10^{-2}\mathrm{i}$ & ~~$0.9895+4.892\times 10^{-3}\mathrm{i}$~ \\
       $\beta=e^\mu-1$  & $0.9779+1.033\times 10^{-2}\mathrm{i}$ & ~~$0.9895+4.890\times 10^{-3}\mathrm{i}$~ \\
       asymptotic solution \quad & $0.9805 +0.933\times 10^{-2}\mathrm{i}$& ~~$0.9902+4.666\times 10^{-3}\mathrm{i}$~ \\
  \end{tabular}
  \caption{Numerical solutions for the eigenvalue $c=c_\mathrm{r}+\mathrm{i}c_\mathrm{i}$ for $m=1$,  $\Omega=1-s\mu^2$,  $s=0.12$ and 0.06,  and three profiles of $\beta$ with $\beta=0$ and $\beta'=1$ at $\mu_\star=0$. The solutions are computed by a shooting method. The three profiles have the same asymptotic prediction for $c$, given by (\ref{4.21}) and shown in the last row of the table.   }
  \label{T1}
  \end{center}
\end{table}

To proceed further, we rewrite equation (\ref{4.21}) as
\begin{equation}\label{4.22}
  \frac{c-\Omega_\star}{(c-\Omega_\star)I_-+J_-}+\frac{c-\Omega_\star}{(c-\Omega_\star)I_++J_+}=\mathrm{i}Q,
\end{equation}
where
\refstepcounter{equation}
$$
  I_-=\int_{-1}^{\mu_\star} (1-\mu^2)\,\mathrm{d}\mu=\mu_\star-\tfrac{1}{3}\mu_\star^3+\tfrac{2}{3},\quad I_+=\int_{\mu_\star}^1(1-\mu^2)\,\mathrm{d}\mu=-\mu_\star+\tfrac{1}{3}\mu_\star^3+\tfrac{2}{3},
$$
$$
  J_-=\int_{-1}^{\mu_\star}(\Omega_\star-\Omega)(1-\mu^2)\,\mathrm{d}\mu,\quad J_+=\int_{\mu_\star}^1(\Omega_\star-\Omega)(1-\mu^2)\,\mathrm{d}\mu, \quad  Q=\frac{2\pi \Omega_\star}{|\beta_\star'|(1-\mu_\star^2)^2}\, . \eqno{(\theequation a,b,c,d,e)} \label{4.23}
$$
In general, equation (\ref{4.22}) is a quadratic equation for $c$ and its solution is
\begin{align}\label{4.24}
  c=\,&\Omega_\star+\frac{1}  {2(I-\mathrm{i}QI_+I_-)}\Big\{-J+\mathrm{i}Q(I_-J_++I_+J_-)   \nonumber \\
& \quad \pm \left[J^2-Q^2(I_+J_--I_-J_+)^2+2\mathrm{i}Q(I_+J_--I_-J_+)(J_+-J_-)\right]^{{1}/{2}}\Big\},
\end{align}
where
\begin{equation}\label{4.25}
  I=I_++I_-=\tfrac{4}{3}, \quad J=J_++J_-=\int_{-1}^1(\Omega_\star-\Omega)(1-\mu^2)\, \mathrm{d}\mu.
\end{equation}
Of the two solutions given by (\ref{4.24}), only those with $c_\mathrm{i}>0$ are valid. At this point, it is not straightforward to obtain an exact condition for $c_\mathrm{i}>0$ to hold, but it is easy to find a sufficient condition as follows. Given that the second line of (\ref{4.24}) has both positive and negative signs, if its first line already has a positive imaginary part, then at least one of the solutions has positive $c_\mathrm{i}$. Therefore, a sufficient condition for instability is
 \begin{equation}\label{4.26}
 \ImIm \left(\frac{\mathrm{i}Q(I_-J_++I_+J_-)-J}{2(I-\mathrm{i}QI_+I_-)} \right)=\frac{\pi\Omega_\star(I_-^2J_++I_+^2J_-)}{|\beta_\star'|(1-\mu_\star^2)^2\left(I^2+Q^2I_+^2I_-^2\right)}>0.
 \end{equation}
A simple example is the situation where $|\Omega_\star|$ is the maximum of $|\Omega|$, then according to (\ref{4.23}$c,d$), both $\Omega_\star J_+$ and $\Omega_\star J_-$ are positive, and (\ref{4.26}) is guaranteed. This leads to the conclusion: \textit{if the angular velocity of the zonal flow is greatest at the critical level, then the flow is unstable.} This agrees with the statement earlier that the semicircle rules always allow instability for such flows, discussed at the end of \S 3. For model solar differential rotation profiles, $\Omega$ is indeed largest at the equator, and so provided $\beta$ passes (transversely) through zero there, the flow is always unstable. The instability induced by the magnetic profile  $\beta=\sigma\mu$ that we showed in figures \ref{F0} and \ref{F2} belongs to this category.


\subsection{ The solution for $\mu_\star=0$ and even $\Omega$} \label{S5.2}


The solutions (\ref{4.24}) may be further simplified and yield transparent results when the critical level is located at the equator ($\mu_\star=0$) and $\Omega$ has even or odd symmetry, which we discuss in \S \ref{S5.2} and \S \ref{S5.3}, respectively.

\begin{figure}
  \centering
  \includegraphics[width=0.495\linewidth]{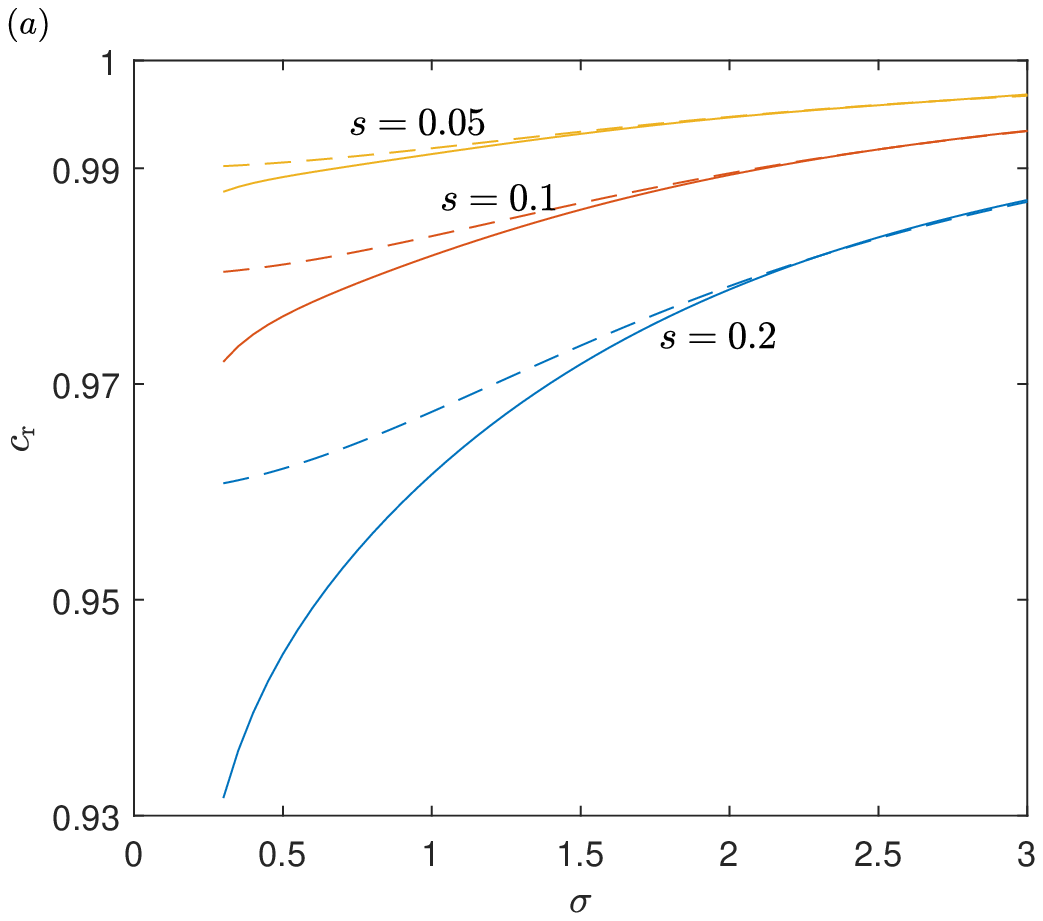}
  \includegraphics[width=0.495\linewidth]{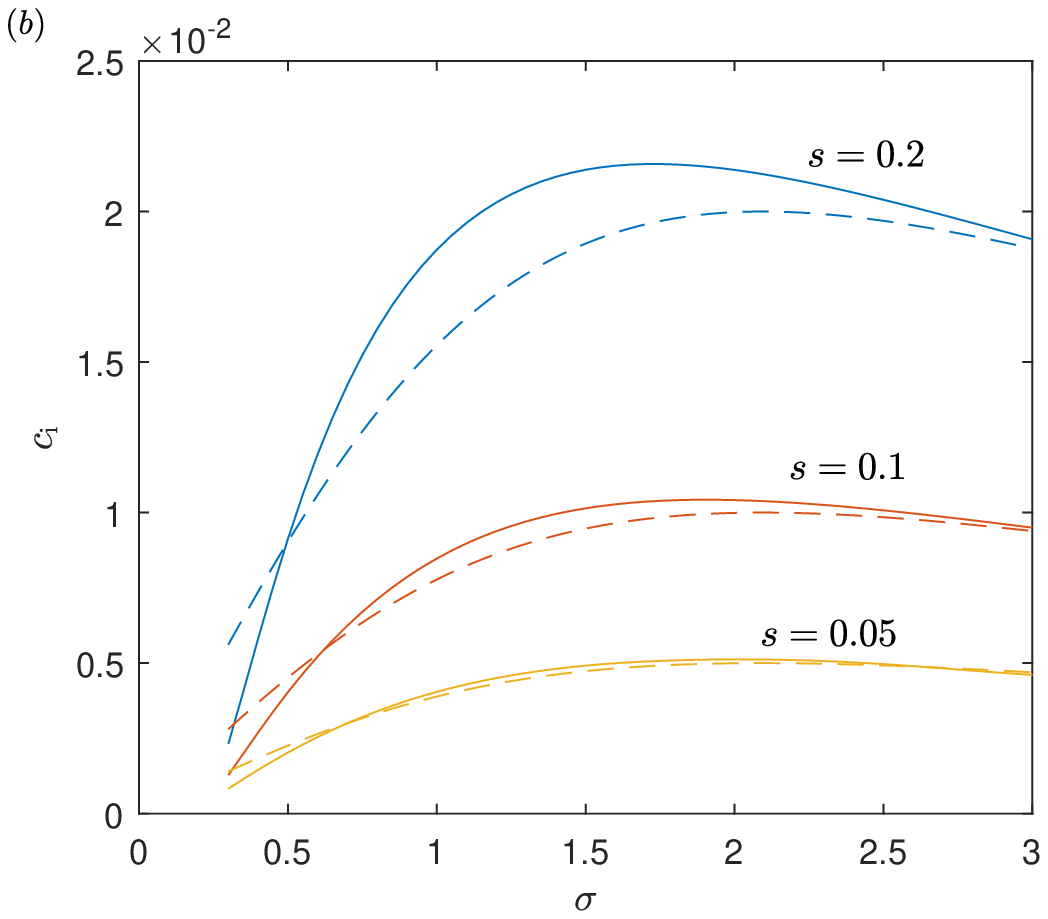}
  \caption{Solutions of $c=c_\mathrm{r}+\mathrm{i}c_\mathrm{i}$ versus $\sigma$ for the profiles $\Omega=r-s\mu^2$ and $\beta=\sigma\mu$ with $r=1$, $s=0.05,0.1,0.2$ and wavenumber $m=1$.  The asymptotic solution (\ref{4.30}) 
  is plotted by dashed lines, and the  numerical solutions are plotted by solid lines. 
  }\label{F3}
\end{figure}

The case in which $\mu_\star=0$ and $\Omega$ is an even function of $\mu$  is  perhaps most relevant to the Sun,  and  therefore, most studies on the clamshell instability focus on this case (for example, \citet{Gilman97, Cally01, Cally03, Miesch07sustained,Miesch07}).  We then have $J_+=J_-$ and $I_+=I_-=2/3$ and the two solutions of (\ref{4.24}) are
\refstepcounter{equation}\label{4.27}
$$
  c=\Omega_\star+\frac{3\mathrm{i}Q J_+}{6-2\mathrm{i}Q}\quad \mathrm{and} \quad c=\Omega_\star-\tfrac{3}{2} J_+. \eqno{(\theequation a,b)}
$$
The first solution (\ref{4.27}$a$) is complex, and may give an unstable mode.  Using the original variables, (\ref{4.27}$a$) becomes
\begin{equation}\label{4.28}
  c=\Omega_\star - \frac{3\pi \Omega_\star\displaystyle\int_0^1 (\Omega_\star-\Omega)(1-\mu^2)\,\mathrm{d}\mu}{2\pi\Omega_\star +3\mathrm{i}|\beta_\star'|}\, .
\end{equation}
Its imaginary part is
\begin{equation}\label{4.29}
 c_\mathrm{i}=\frac{9\pi\Omega_\star|\beta_\star'|\displaystyle\int_0^1(\Omega_\star-\Omega)(1-\mu^2)\,\mathrm{d}\mu}{4\pi^2\Omega_\star^2+9{\beta_\star'}^2}\, .
\end{equation}
For the standard profile 
\begin{equation}\label{4.30a}
  \Omega=r-s\mu^2,\quad  \beta=\sigma\mu,
\end{equation}
we find
\refstepcounter{equation}\label{4.30}
$$
c_\mathrm{r}=r-\frac{4\pi^2r^2s}{20\pi^2r^2+45\sigma^2}\, ,\quad c_\mathrm{i}=\frac{6\pi|\sigma|rs}{20\pi^2 r^2+45 \sigma^2}\, . \eqno{(\theequation a,b)}
$$
As discussed in the last paragraph of \S \ref{4.2.3}, this solution is accurate in the limit when $s$ is small or $|\sigma|$ is large. Note that large $|\sigma|$ at fixed $r$ and $s$ corresponds to the situation where $\Omega$ as a whole is weak compared to $\beta$. The detailed analysis for this case is shown in appendix A.  The results of (\ref{4.30}) are plotted in figure \ref{F3} by dashed lines, and may be compared to the numerical solutions shown by solid lines. We see that the asymptotic solution gives very good predictions in general, and that these become more precise as $s$ decreases or as $\sigma$ increases.

With the asymptotic solution for the growth rate $c_\mathrm{i}$  given by (\ref{4.30}$b$), we may address the question of whether instability persists when the parameters approach limiting values. \citet{Gilman97} raised the question of whether there is a lower limit of positive $s$ and an upper limit of $\sigma$, for the instability to take place. These thresholds do not seem to exist according to their numerical solutions, but the unstable mode becomes more and more singular at the critical level as $s$ decreases or $\sigma$ increases, causing numerical difficulties. Our asymptotic solution can easily address this problem: (\ref{4.30}$b$) indicates that such limits indeed do not exist: instead, as  $s\rightarrow 0^+$ or  $\sigma\rightarrow\infty$, $c_\mathrm{i}$ remains positive at $O(s)$  or $O(\sigma^{-1})$ provided that $rs>0$.  
The clamshell instability is therefore quite different from hydrodynamic shear instability on a sphere, which requires the shear to exceed a threshold ($s/r>0.29$, \citet{Watson81}) to overcome the stabilising effect of the rotation.
 The fact that the instability survives for arbitrarily strong magnetic field is also surprising, but we note that regardless of the strength of $\sigma$, the magnetic field always vanishes at $\mu_\star=0$, and it is this feature that plays a fundamental role in inducing the instability.

The solution (\ref{4.29}) can also provide insights into the instability for general flow profiles, not only those related to solar differential rotation. From  the condition of $c_\mathrm{i}>0$, we have:  \textit{for $\mu_\star=0$ and an even profile of $\Omega(\mu)$, the condition for instability is}
\begin{equation}\label{4.31}
  \Omega_\star\int_{0}^1(\Omega_\star-\Omega)(1-\mu^2)\,\mathrm{d}\mu>0.
\end{equation}
This indicates that the flow is prone to instability when the angular velocity $\Omega_\star$ at the critical level (the equator in this case) is large compared to $\Omega(\mu)$ on the rest of the sphere. If $|\Omega_\star|$ is the largest among all $|\Omega|$, then the flow is definitely unstable. Interestingly,  (\ref{4.31}) only involves the hydrodynamic shear,  and $\beta$ does not affect this condition once $\mu_\star=0$ is set.  Equation  (\ref{4.31}) is also quite different from conditions for hydrodynamic shear instability: the latter usually involve constraints on the curvature of the basic-flow profile (cf.\ Rayleigh's inflection-point theorem),  but (\ref{4.31}) does not involve $\Omega''$ at all.



We can also derive a bound for the growth rate from (\ref{4.29}), namely
\begin{equation}\label{4.32}
  c_\mathrm{i}< \frac{9\pi |\Omega_\star||\beta_\star'|\max|\Omega_\star-\Omega|\displaystyle\int_0^1(1-\mu^2)\,\mathrm{d}\mu}{12\pi|\Omega_\star\beta_\star'|}\leq \tfrac{1}{2}\max |\Omega_\star-\Omega|\,,
\end{equation}
%
%
where the inequality $x^2 + y^2 \geq 2|xy|$ has been used in the denominator. Again, once $\mu_\star=0$ is set by the magnetic field, this bound only involves the hydrodynamic shear. We note that the semicircle rules studied in \S 3 suggest that the magnetic field may increase the radii of the semicircles and hence the bound for the growth rate, but this does not happen in (\ref{4.32}). However, we also note that the semicircle rules apply to general velocity and field profiles, and (\ref{4.32}) is the result for the more specific situation in which the rotation profile is even with weak shear, and the magnetic profile passes through zero at one location.
%

Clearly when $\Omega$ is even, $\Omega_1$ is also even, and from (\ref{4.16}) we have $A_-=-A_+$. Thus the critical layer makes the tilting modes opposite on the two hemispheres (as also shown in figure \ref{F2}$c$), which explains the typical clamshell pattern shown in figure \ref{F0}.

Finally, we comment on the other solution (\ref{4.27}$b$). The physical meaning of this solution is that it makes  the singularity of $H_1'$ given in (\ref{4.9}) vanish. Thus to leading order, the weak shear does not trigger the singularity of the critical level of the tilting mode.  One may need to go to higher orders in the asymptotic expansion, which may contain  potential singularities and yield a even smaller $c_\mathrm{i}$.  Our numerical solution suggests that for the basic state profiles (\ref{4.30a}), solution (\ref{4.27}$b$) corresponds to a quasi-mode instead of a normal mode.  However, when the critical level $\mu_\star$ is slightly off the equator, (\ref{4.27}$b$) becomes an unstable normal mode, as we will show in \S \ref{S5.4}.


\subsection{The solution for $\mu_\star=0$ and odd $\Omega-\Omega_\star$} \label{S5.3}

\begin{figure}
  \centering
  \includegraphics[width=0.49\linewidth]{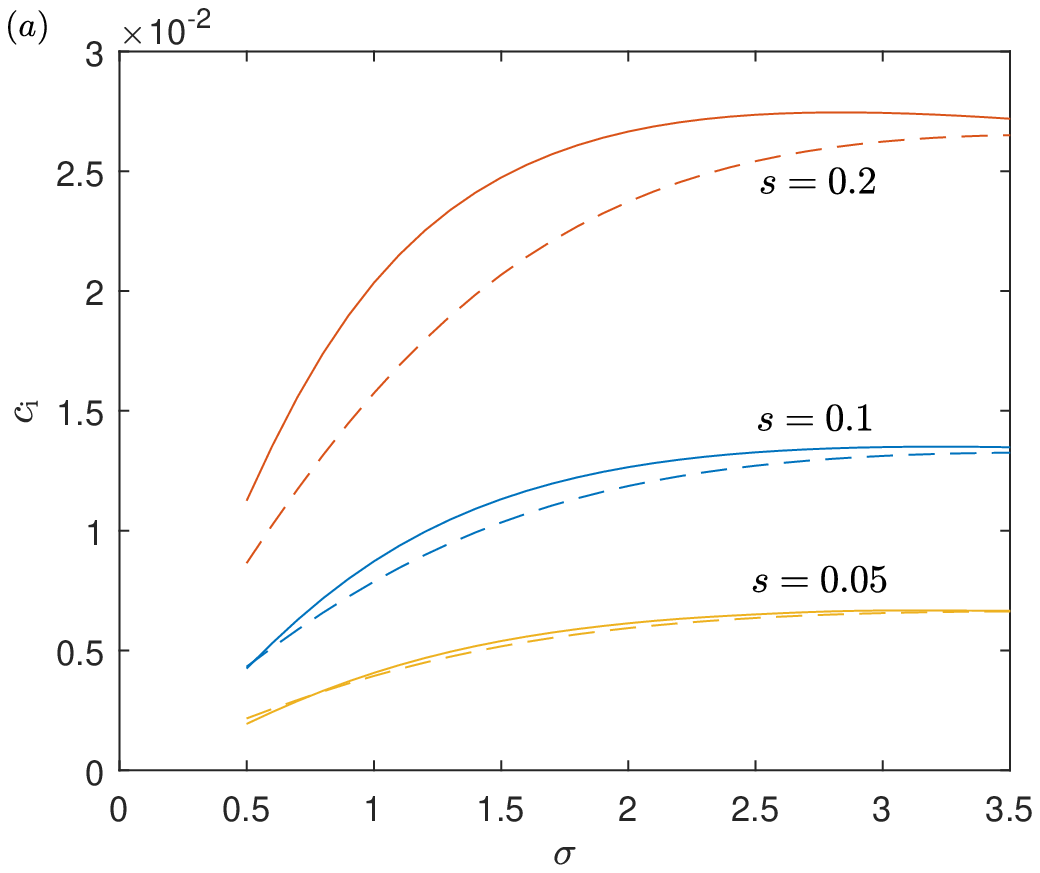}
  \includegraphics[width=0.49\linewidth]{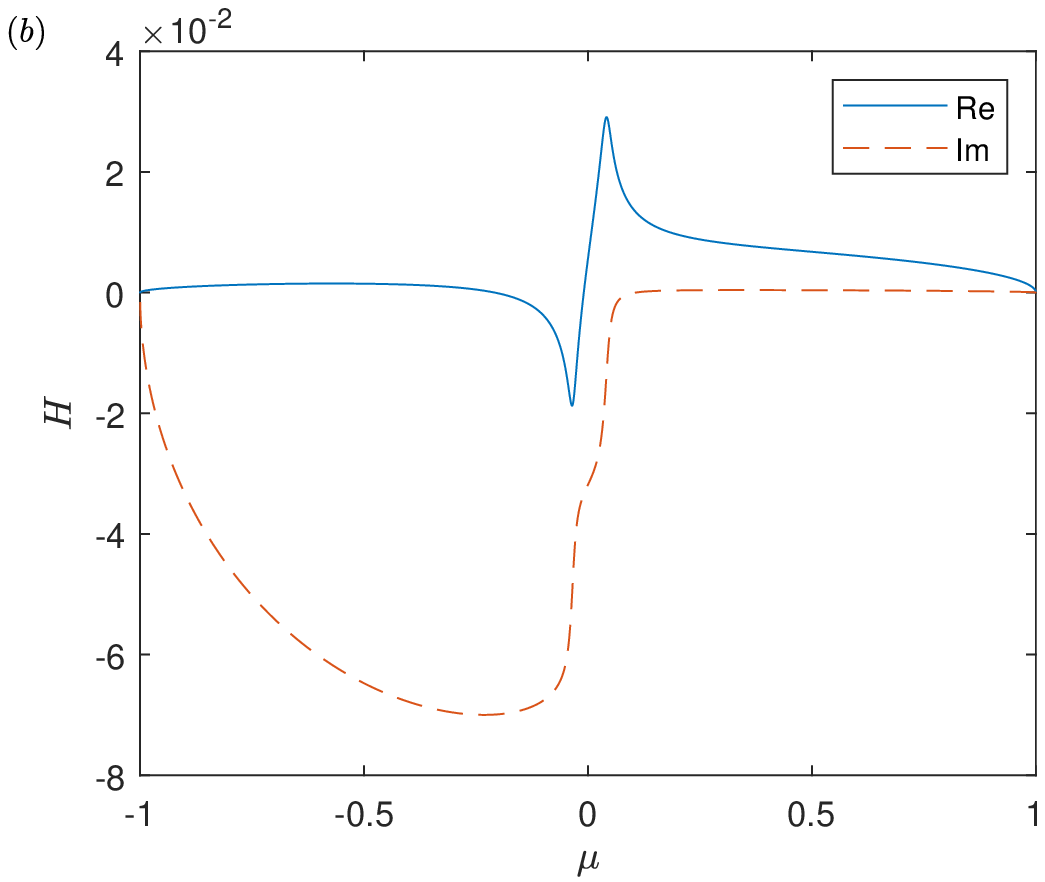}
  \caption{Instability of the profiles $\Omega=r+s\mu$ and $\beta=\sigma \mu$ with $r=1$,  $m=1$.  (a) Curves of $c_\mathrm{i}$ versus $\sigma$ for $s=0.05,0.1,0.2$.  The asymptotic solution (\ref{4.33}) with (\ref{4.36}) is plotted by dashed lines, and the numerical solution is plotted by solid lines. 
  (b) Numerical solution showing $H$ for $r=\sigma=1$ and $s=0.1$, with eigenvalue $c=0.96+0.0087\mathrm{i}$. 
  }\label{F4}
\end{figure}

The situation in which $\mu_\star=0$ and the shear profile $\Omega-\Omega_\star$ is an odd function of $\mu$ is less relevant to the Sun, but as a basic model it is still of interest to fluid mechanics and we may draw useful general conclusions in this case. Here we have $J_+=-J_-$, $I_-=I_+=2/3$  and the imaginary part of (\ref{4.24}) can be simplified to
\begin{equation}\label{4.33}
  c_\mathrm{i}=\pm\sqrt{\frac{81|Q|J_+^2}{8(Q^2+9)\bigl(\sqrt{Q^2+9}+|Q|\bigr)}}.
\end{equation}
%
Except for the special case of $J_+=0$ (for which we would need to pursue higher orders of the asymptotic expansion),  there is always a positive $c_\mathrm{i}$ solution and so, surprisingly, we may conclude that \textit{if $\mu_\star=0$ and $\Omega(\mu)-\Omega_\star$ is odd, the flow is  always unstable. } 
Bounding the denominator of (\ref{4.33}) from below via $(Q^2+9)(\sqrt{Q^2+9}+|Q|) > 9 \times 2 |Q|$, we obtain a bound for the unstable growth rate:
\begin{equation}\label{4.34}
  c_\mathrm{i}< \tfrac{3}{4}|J_+| \le\tfrac{1}{2}\max |\Omega_\star-\Omega|.
\end{equation}
Interestingly, this bound is the same as (\ref{4.32}), but we expect it to be looser since we have bounded  positive terms by zero. To give a concrete example for this instability, we consider
\begin{equation}\label{4.35}
  \Omega=r+s\mu,\quad \beta=\sigma\mu\,,
\end{equation}
where $\Omega$ features a `linear shear' profile, analogous to Couette flow. Then we have
\begin{equation}\label{4.36}
  Q=\frac{2\pi r}{|\sigma|}\, , \quad J_+=-\frac{s}{4}\, .
\end{equation}
The results of (\ref{4.33}) with (\ref{4.36}) are shown figure \ref{F4}$(a)$, where they are compared to the numerical solution. Again good agreement is found and the agreement improves as $s$ becomes smaller or $\sigma$ becomes larger. The behaviour of $c_\mathrm{i}$ is similar to the previous case of even $\Omega$, as is $c_\mathrm{r}$ (not shown). An example for the solution of $H$ is shown in figure \ref{F4}$(b)$. Because $\Omega$ as a whole is neither even nor odd, $H$ has no symmetry property either.

\subsection{An example for $\mu_\star\neq 0$} \label{S5.4}

When the critical level is off the equator (i.e.~$\mu_\star\neq0$), there is no obvious symmetry property that can simplify the asymptotic solution for $c$ given by (\ref{4.24}). We have not been able to obtain general conclusions regarding the condition for instability in this case, but the asymptotic solution can still be helpful in understanding numerical results. As an example, we consider
\begin{equation}\label{4.38}
  \Omega=1-0.1\mu^2,\quad \beta=\mu-d.
\end{equation}
The zonal flow features the solar differential rotation as before, and $\beta$ is a linear profile with the critical level located at $\mu_\star =d$. When $d=0$ we recover the standard configuration (\ref{4.30a}).

\begin{figure}
  \centering

  \includegraphics[width=0.49\linewidth]{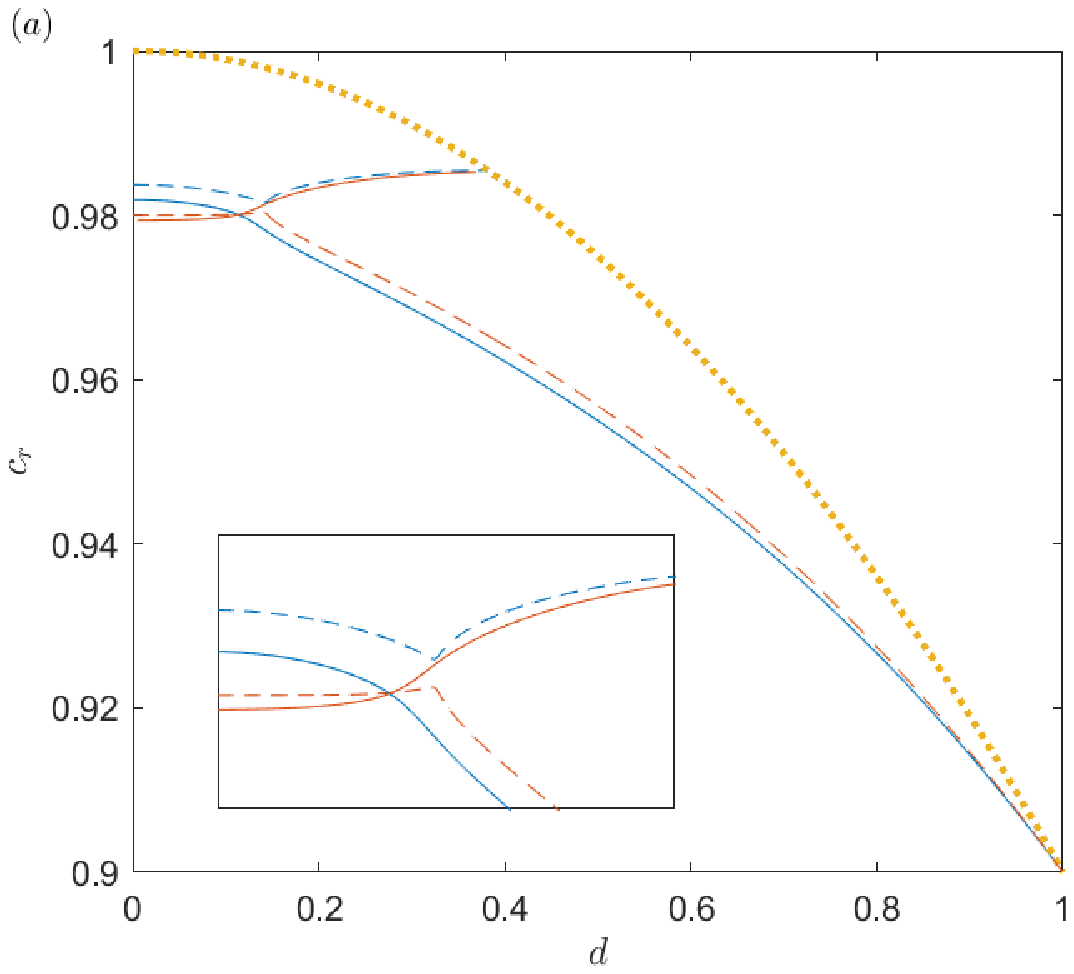}
  \includegraphics[width=0.49\linewidth]{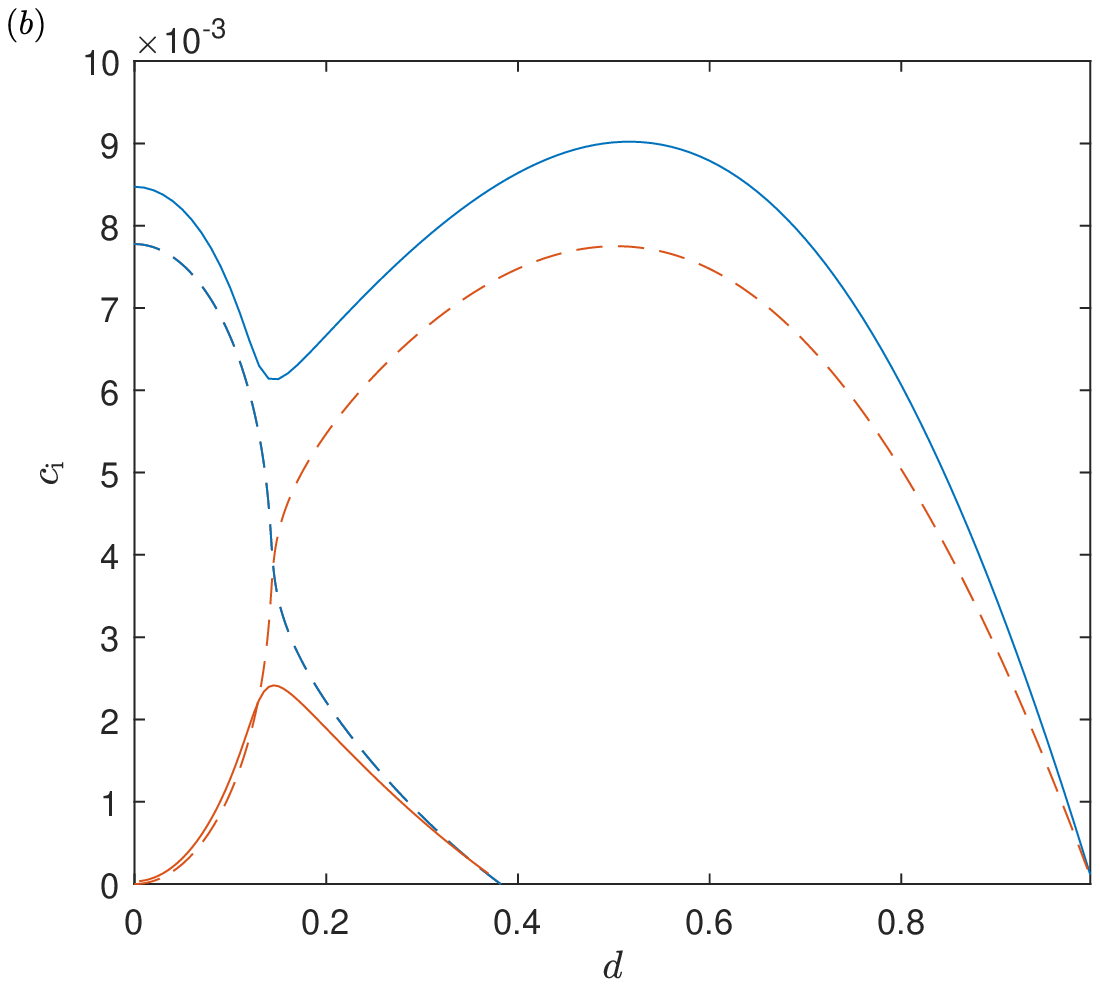}
  \caption{Eigenvalue $c=c_\mathrm{r}+\mathrm{i}c_\mathrm{i}$ versus $d$ when $\Omega=1-0.1\mu^2$, $\beta=\mu-d$ and $m=1$.   The solid lines represent the numerical solution, and the dashed lines are the results of the asymptotic solution (\ref{4.24}).   The dotted line in panel ($a$) represents the rotation rate $\Omega$ at the critical level $\mu_\star=d$, i.e.  $\Omega_\star=1-0.1d^2$. The inset in panel ($a$) shows the region where the two eigenvalues become close.
  }\label{F5}
\end{figure}

The asymptotic solution for $c$ computed from (\ref{4.24}) versus $d$ is shown in figure \ref{F5} (dashed lines), where it is compared to the numerical solution (solid lines).  In figure \ref{F5}$(a)$,  we also plot the value of $\Omega$ at the critical level,  $\Omega_\star=1-0.1 d^2$ (dotted line).   The main feature of this asymmetric case is that  both solutions of (\ref{4.24}) can have positive $c_\mathrm{i}$,  and thus there are two branches of unstable modes. When $d$ slightly departs from zero, the mode that corresponds to (\ref{4.27}$b$) becomes unstable (red dashed line), and its growth rate dominates over the other unstable mode for a large range of $d$. On the other hand, as $d$ increases, the unstable mode that corresponds to (\ref{4.27}$a$) (blue dashed line) is weakened significantly, and disappears at $d\approx 0.37$. Again good agreement is found between the asymptotic and numerical results,  but interestingly,  there is a topological difference between them:   the asymptotic solution predicts that when the two eigenvalues are close, the curves of $c_\mathrm{i}$ intersect while those of $c_\mathrm{r}$ avoid the intersection, while the opposite is true of the numerical solution.

It appears from figure \ref{F5}$(a)$ that $c_\mathrm{r}$ is always smaller than $\Omega_\star$, which demands an explanation. When the mode of the blue dashed line (or the red solid line) has $c_\mathrm{r}$ approach $\Omega_\star$ at $d\approx 0.37$ (left panel), the corresponding  $c_\mathrm{i}$ (right panel) approaches zero. 
Thus,  $\Omega_\star$ appears to be an upper bound for $c_\mathrm{r}$ for unstable modes. There is an underlying reason for this phenomenon, related to the conservation of angular momentum, as we will explain in \S\ref{angular_momentum}.

\begin{figure}
  \centering
  \includegraphics[width=0.495\linewidth]{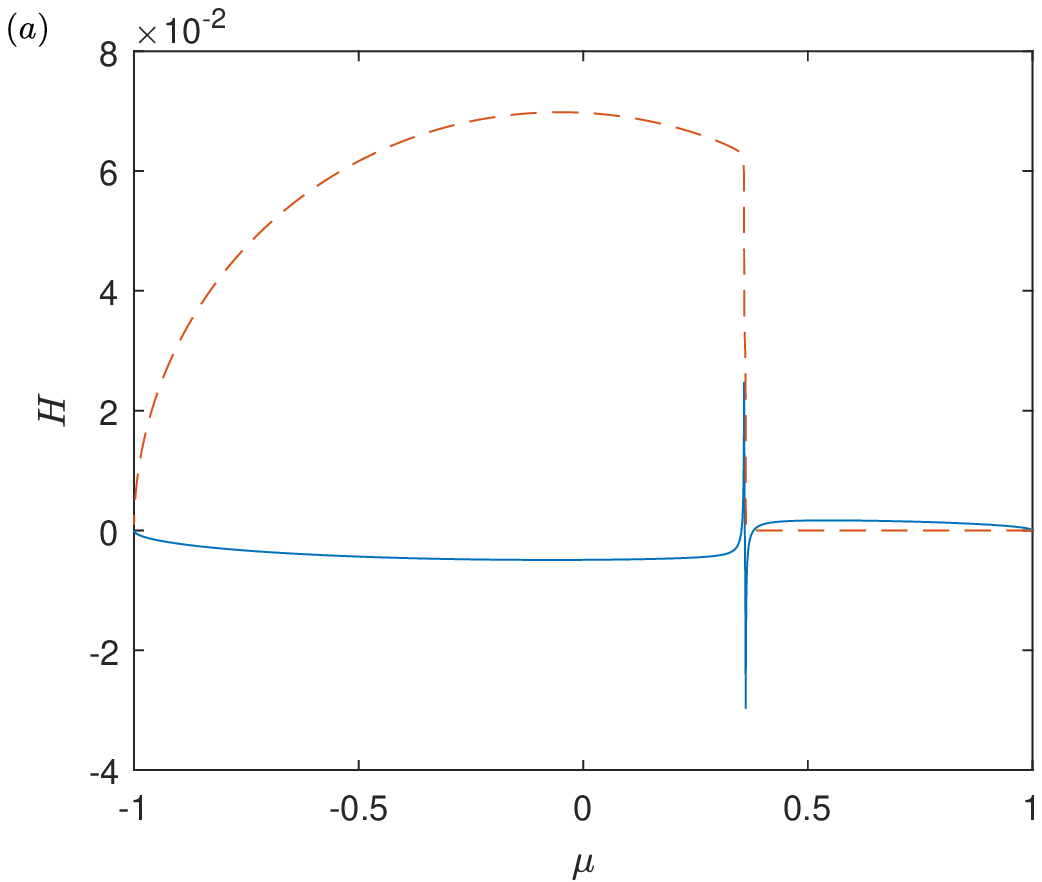}
  \includegraphics[width=0.495\linewidth]{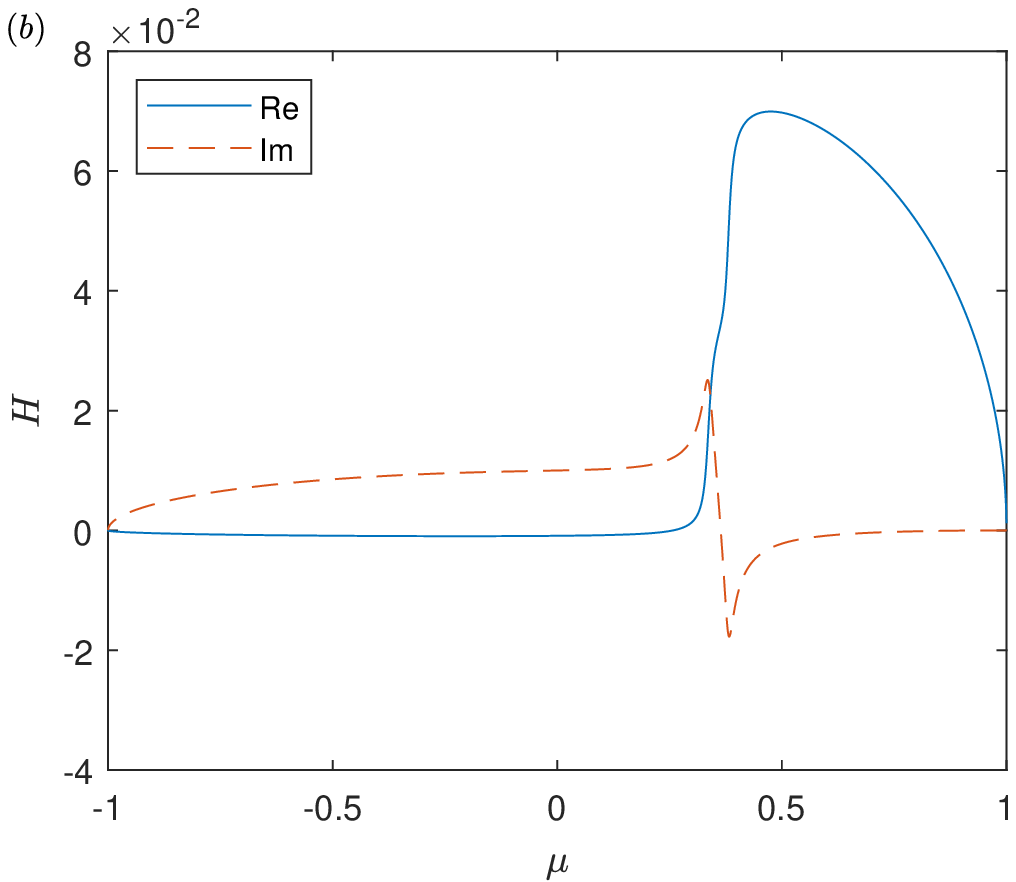}
  \caption{Two numerical solutions of the eigenfunction $H$ corresponding to figure \ref{F5} at $d=0.36$.  The eigenvalues are $(a)$ $c=0.985+2.05\times 10^{-4}\mathrm{i}$, $(b)$ $c=0.965+8.36\times 10 ^{-3}\mathrm{i}$.  The case of panel $(a)$ is the one with almost the smallest growth rate that we can compute.
  }\label{F6}
\end{figure}

The eigenfunctions of the two unstable modes at $d=0.36$ are plotted in figure \ref{F6}.  Figure \ref{F6}$(a)$ is the mode with the smaller growth rate $c_\mathrm{i}$. In fact,  we have chosen $d$ such that this mode is almost as close to a neutral mode as we can compute numerically.  The very small $c_\mathrm{i}$ makes the critical layer have a very fine structure.  There is a significant difference in the amplitudes of the tilting modes on the two sides of the critical layer.  Figure \ref{F6}$(a)$ has a much larger, stronger tilting mode to the left of the critical level,  while figure \ref{F6}$(b)$ has the opposite feature.

\subsection{The conservation of angular momentum} \label{S5.5}
\label{angular_momentum}
The asymptotic analysis clearly indicates that the critical layer plays a fundamental role in making the flow unstable.  In our previous studies of instability induced by critical layers \citep{Riedinger,Wang2018,Wang22},  conservation of momentum provides a useful tool for understanding the mechanism of the instability. Indeed, it has been found that the critical layer provides a source of mean-flow momentum, which drives the exponential growth of the outer flow.
In the current problem in spherical geometry, the relevant conservation law is that of angular momentum (\ref{2.33}), namely,
\begin{equation}\label{4.39a}
  \int_0^\pi 2\pi \sin^2\theta\, \frac{\partial \Delta U}{\partial t}\,\mathrm{d}\theta=0.
\end{equation}
It is of interest to understand how this conservation is achieved,  i.e.~how different regions contribute to the integral and balance each other.

Substituting (\ref{2.20}),  (\ref{2.24a}) and (\ref{2.26}) into (\ref{2.32}),   we can derive the rate of change of angular momentum per latitude:
\begin{equation}\label{4.39}
2\pi\sin^2\theta\, \frac{\partial \Delta U}{\partial t}= \frac{\partial}{\partial \theta}\frac{\partial L}{\partial t},
\end{equation}
with
\begin{equation}\label{4.40}
  \frac{\partial L}{\partial t}=\Bigl[4\pi(1-\mu^2)\left\{ \bigl[|\Omega-c|^2-\beta^2\bigr] \ImIm(H{H'}^*)-c_\mathrm{i}\Omega' |H|^2\right\}\Bigr] e^{2c_\mathrm{i} t}.
\end{equation}
%
Here $L(\theta,t)$ represents the  total mean-flow angular momentum between the north pole and co-latitude  $\theta$. Recall that the primes denote derivatives with respect to $\mu=\cos \theta$.  For the clamshell instability studied above, $\Omega-c$, $\Omega'$ and $c_\mathrm{i}$ are all small at order $O(\varepsilon)$, so that outside, or at the edge of the critical layer, we have
\begin{equation}
 \frac{\partial L}{\partial t}=-4\pi(1-{\mu}^2)\beta^2 \ImIm(H{H'}^*)\, e^{2c_\mathrm{i}t},
\end{equation}
to leading order of $\varepsilon$. This result corresponds to the fact that the Maxwell stress, i.e. $\overline{a_\ell b_\ell}$ in (\ref{2.32}) has the dominant contribution to the mean-flow response, whilst the Reynolds stress $\overline{u_\ell v_\ell}$ has a minor effect due to the weak shear. We can then study the integral of (\ref{4.39}) over $\theta$ in different regions. We define $\theta_\star$ as the value of $\theta$ at the critical level (i.e. $\cos\theta_\star=\mu_\star$) and set $\Delta=O(\varepsilon^{{1}/{2}})$ as the half-thickness of the critical layer. Then, outside the critical layer the integrals are
\refstepcounter{equation}
$$
   \int_0^{\theta_\star-\Delta}2\pi\sin^2\theta \, \frac{\partial \Delta U}{\partial t}\, \mathrm{d}\theta=\frac{\partial L}{\partial t}\,\bigg|_{\mu=\mu_\star+\Delta\,{\sin\theta_\star}}=8\pi c_\mathrm{i}\Omega_0|A_+|^2e^{2c_\mathrm{i}t}I_+, \eqno{(\theequation a)}
$$
$$
  \int_{\theta_\star+\Delta}^\pi 2\pi\sin^2\theta \, \frac{\partial \Delta U}{\partial t}\, \mathrm{d}\theta=-\frac{\partial L}{\partial t}\, \bigg|_{\mu=\mu_\star-\Delta\,{\sin\theta_\star}}=8\pi c_\mathrm{i}\Omega_0|A_-|^2e^{2c_\mathrm{i}t}I_-, \eqno{(\theequation b)} \label{4.42}
$$
where $I_-$ and $I_+$ are the positive quantities defined in (\ref{4.23}) and we have used $H=H_0$ and $H'=\varepsilon H_1'$ from \S 4.2.1 as the leading-order approximation on the edge of the critical layer. Inside the critical layer, using the inner solution given in \S 4.2.2, we find the integral to be
\begin{equation}\label{4.44}
  \int_{\theta_\star-\Delta}^{\theta_\star+\Delta}2\pi\sin^2\theta\, \frac{\partial \Delta U}{\partial t}\, \mathrm{d}\theta=-\frac{\partial L}{\partial t}\, \bigg|_{\eta\rightarrow -\infty}^{\infty}=\frac{4\varepsilon\pi^2(1-\mu_\star^2)|\alpha_1|^2}{|\beta_\star'|}\ReRe \left(\frac{1}{c_1-\Omega_{1\star}}\right)e^{2c_\mathrm{i}t}.
\end{equation}
Applying the relations between the constants $\alpha_1$, $\alpha_2$, $A_-$ and $A_+$ given in (\ref{4.16}) and (\ref{4.19}), we can show that the value of (\ref{4.44}) exactly cancels the sum of (\ref{4.42}$a$) and (\ref{4.42}$b$), 
and results in  (\ref{4.39a}) being satisfied. The critical layer thus  provides a source of angular momentum which balances that of the outer flow.  We note that without the critical-layer angular momentum (\ref{4.44}),  the only possibility that (\ref{4.42}$a$) and (\ref{4.42}$b$) could add up to zero is if $c_\mathrm{i}=0$,  i.e.~the tilting modes by themselves have to be  neutral modes.  This demonstrates how the angular momentum provided by the critical layer is necessary to drive the instability.

We may gain some further insights by considering the sign of the mean angular momentum inside and outside the critical layer. Without loss of generality, we consider the case in which $\Omega_0>0$. Then both (\ref{4.42}$a$) and (\ref{4.42}$b$),  giving the mean angular momentum of the tilting components, are positive, so that the contribution from the critical layer (\ref{4.44}) must be negative to make the conservation law (\ref{4.39a}) possible.  Since
\begin{equation}\label{4.45}
  \ReRe \left(\frac{1}{c_1-\Omega_{1\star}}\right)=\frac{c_{1\mathrm{r}}-\Omega_{1\star}}{(c_{1\mathrm{r}}-\Omega_{1\star})^2+c_\mathrm{1i}^2},
\end{equation}
we require
\begin{equation}
c_{1\mathrm{r}}<\Omega_{1\star}\quad  \mathrm{or} \quad c_\mathrm{r}<\Omega_\star.  \label{4.45}
\end{equation}
This means that \textit{for any unstable mode, the real part of the phase velocity must be smaller than the  velocity of the zonal flow at the critical level.}  We can verify that all of the solutions we have showed previously satisfy this condition. For example,  in (\ref{4.30}$a$) we have $c_\mathrm{r}<r$ for $s>0$,  and in figure \ref{F5}$(a)$ the curve of  $\Omega_\star$ is  always above that of $c_\mathrm{r}$. In the situation of figure \ref{F5}$(a)$, we may also view (\ref{4.45}) as a necessary condition for the existence of an unstable mode: when $c_\mathrm{r}$ is about the exceed $\Omega_\star$ at $d\approx 0.37$, the unstable mode disappears.


We may undertake a similar analysis for the conservation of mean toroidal field as shown by (\ref{2.34}), but we were not able to obtain straightforward general conclusions. This is mainly because the local integral of $\partial_t \Delta A$ in the critical layer has a less transparent expression. Nevertheless, we document these results in appendix B for the readers' interest.

\section{Conclusions}\label{S6}
We have studied the linear instability of 2D MHD flows on a sphere, a  problem with potential
application to the instability of the solar tachocline. We  derived  semicircle rules for the complex phase velocity, which provide rigorous bounds for general flow and field profiles. The  terms arising purely from the spherical geometry bring new features to the problem. We  used two bounding methods, which provide two versions of  the semicircle rules, each of which may be tighter for certain types of flows. We   also found that the magnetic field may increase the radii of the semicircles, which does not happen in the case of Cartesian geometry \citep{Hughes01}.


We then undertook an analytical study of the `clamshell instability'. Previous studies have found that the instability tilts the basic magnetic field lines on the two hemispheres in opposite directions, giving a pattern of an opening clamshell \citep{Cally01,Cally03}. We studied this instability theoretically through an asymptotic analysis in the limit of weak shear of the basic zonal flow.  
We  found that if the basic zonal flow is a pure solid body rotation, there exists an eigenmode that slightly tilts the entire magnetic field and makes it  rotate with the zonal flow. We refer to this disturbance as a `tilting mode'. Including an additional weak shear in the zonal flow excites the critical level of the tilting mode, located at the node of the sheared field profile.
 Disturbances exhibit strong singular behaviour near the critical level, inside the critical layer. We  found that the critical layer reverses the direction of tilting and makes the flow unstable.


Through matching the tilting mode and the critical layer, we  derived the asymptotic solution for the complex  phase velocity, from which we obtained properties of the instability for general profiles. Our investigations indicate that the magnetic field only affects the instability through the location of the critical level and its gradient at the critical level; the other details of the field profile do not matter. A sufficient condition for instability is that the critical level is located where the angular velocity of the zonal flow is greatest.  When the zonal flow is even about the equator and the critical level is on the equator, we derived a simple expression for the unstable growth rate, which indicates that the flow is susceptible to instability when the angular velocity at the critical level is large compared to that on the rest of the sphere.  When  the shear of the zonal flow is odd and the critical level is on the equator, the flow is  always unstable. A simple bound for the unstable growth rate  was derived  for these two types of flows with even or odd symmetry properties. In the absence of symmetry, when the critical level is off the equator, there can be two branches of unstable modes. The results of the asymptotic solution are in good agreement with the numerical solutions.

A mechanism for the instability has been provided via the conservation of angular momentum. The critical layer provides a source of angular momentum, which must be balanced by a corresponding sink for the surrounding tilting mode. In order that the angular momentum of the tilting mode and  critical layer have  opposite signs, the phase velocity of the unstable mode must be smaller than the velocity of the zonal flow at the critical level.

Our study reveals several problems that are left for future research. We found that the magnetic field can increase the radii of the semicircles (over those for the purely hydrodynamic flow), but we have not yet found an unstable mode that resides in this new region. It is interesting to investigate whether it can be found for different flow and field profiles.  In addition, when the magnetic field is relatively strong, the destabilising effect of the field is always associated with an increase of the semicircle radius. It remains an open question as  to whether there is a deeper link between these observations.  The theories of \citet{Thuburn96}, \citet{Sasaki12} and \citet{Deguchi21} which may provide different semicircles could be possible routes to approach this problem.

 The clamshell instability we studied occurs for idealised MHD flows with weak shear and a strong field that vanishes at one location.  There are flows with field-induced instabilities that do not belong to this category. For example, strong shear combined with weak magnetic field \citep{Gilman97, Cally01}, magnetic field profiles with multiple zero points  \citep{Dikpati99}, and narrow bands of magnetic field   \citep{Dikpati99, Cally03}. Dissipation may also be of potential interest: our results indicate that for ideal MHD, the field lines on the two sides of the critical layers are separated, so it would be interesting to explore the details of the reconnection caused by diffusion as seen in the simulation of \citet{Cally01}.  It has also been found that diffusion may destabilise the flow even when the zonal flow has no  shear \citep{Sharif05}. Beyond the incompressible MHD setting, instabilities also arise in shallow-water MHD systems in spherical geometry \citep{Marquez17,Gilman02}. A deeper understanding might be gained by studying whether a similar asymptotic analysis is applicable to these instabilities. We also plan to explore the theory of nonlinear critical layers, to understand the saturation of growing disturbances. 


\section*{Acknowledgments}
\noindent
This work is supported by the EPSRC (grant EP/T023139/1), which is gratefully acknowledged.  We thank the referees for their constructive comments, which have helped clarify our discussion, and for providing further useful references.

\section*{Declaration of interests}
\noindent
The authors report no conflict of interest.

\section*{Data access statement}
\noindent
No data were created or analysed in this study.

\appendix
\section{The asymptotic solution for weak zonal flow}

In this appendix, we consider the matched asymptotic expansion for the situation where the zonal flow $\Omega$ as a whole is weak compared to the magnetic field $\beta$. In this case, we no longer require that $\Omega$ is a solid body rotation at leading order.  As before, only the wavenumber $m=1$ is considered since it is the only wavenumber that admits the tilting mode solution (\ref{4.2}). The requirement for $\beta$ is the same as before, i.e.\ that it passes through zero at $\mu_\star$ with a gradient that is of order of unity or larger.  As we noted previously, the derivation here is slightly different to that presented in the main text but the final equation that determines the eigenvalue, equation (\ref{4.21}), remains the same.

We may regard the weak zonal flow as a perturbation to the tilting mode (\ref{4.2}) at $\Omega=0$, which also perturbs $c$ away from zero:
\begin{equation}
\Omega=\varepsilon \Omega_1(\mu),\quad c=\varepsilon c_1 + \cdots . \label{A1}
\end{equation}
For the outer solution of $H$,  the expansion is
\begin{equation}
H=H_0+\varepsilon^2 H_1 + \cdots .  \label{A2}
\end{equation}
$H_0$ is still expressed by the piecewise tilting mode (\ref{4.6}),  but the next order of (\ref{A2}) is now $\varepsilon^2$,  due to the absence of $O(\varepsilon)$ terms in the coefficients of $H$ in (\ref{2.27}).    Substituting (\ref{A1}) and (\ref{A2}) into (\ref{2.27}),  the $O(\varepsilon^2)$ terms yield an equation for $H_1$:
\begin{align}
&\left[\beta^2(1-\mu^2)H_1'\right]'+\left[2\beta(\mu\beta)'-\frac{\beta^2}{1-\mu^2}\right]H_1 \nonumber \\
&\qquad =\left[(\Omega_1-c_1)^2(1-\mu^2)H_0'\right]'+\left[2(\Omega_1-c_1)(\mu\Omega_1)'-\frac{(\Omega_1-c_1)^2}{1-\mu^2}\right]H_0. \label{A3}
\end{align}
Since $H_0$ satisfies (\ref{2.27}) for $c=0$, $\beta=0$ and  $\Omega=\Omega_1-c_1$ for \emph{any} $\Omega_1$ and $c_1$ (see the exact solution (\ref{4.1})),   we may simplify (\ref{A3}) to
\begin{equation}
\left[\beta^2(1-\mu^2)H_1'\right]'+\left[2\beta(\mu\beta)'-\frac{\beta^2}{1-\mu^2}\right]H_1 =2c_1(\Omega_1-c_1)H_0.\label{A4}
\end{equation}
We may now solve for $H_1$ using the same method as before. We find $\varepsilon^2 H_1\sim \varepsilon^2 (\mu-\mu_\star)^{-1}$, which becomes as large as $H_0$ when $\mu-\mu_\star\sim \varepsilon ^2$. Hence  the critical layer has the small length scale of $\varepsilon^2$.

For the inner solution, and the leading-order terms in the local equation are still those with spatial derivatives, due to the small length scale:
\begin{equation}
(SH')'=0,\quad S=(\Omega-c)^2-\beta^2. \label{5.6}
\end{equation}
In order to obtain a local solution that is uniformly valid throughout the critical layer,  $\Omega-c$ needs to balance $\beta$,  which implies
that $\Omega-c\sim\beta\simeq \beta_\star'(\mu-\mu_\star)= O(\varepsilon^2)$ in the critical layer. But in (\ref{A1}), both $\Omega$ and $c$ are at $O(\varepsilon)$, so the only possibility is that   $c$ is the same as $\Omega$ at order $\varepsilon$, and their difference is at order $\varepsilon^2$, which means
\begin{equation}
 c=\Omega_{\star}+\varepsilon^2 c_2 +\cdots.
 \label{A7}
\end{equation}
Introducing the local coordinate
\begin{equation}
  \eta=\frac{\mu-\mu_\star}{\varepsilon^2}\, ,\qquad H=\mathcal{H}(\eta) + \cdots, \label{A6}
\end{equation}
(\ref{5.6}) becomes 
\begin{equation}\label{A8}
  \frac{\mathrm{d}}{\mathrm{d}\eta}\left[(c_2^2-\beta'^2_\star \eta^2)\, \frac{\mathrm{d}\mathcal{H}}{\mathrm{d}\eta}\right]=0.
\end{equation}

The remainder of the calculation is the same as \S 4. We solve (\ref{A4}) and (\ref{A8}), and then match them to find the equation for the eigenvalue. The final result is
\begin{equation}\label{A9}
  \frac{c_2}{\displaystyle \int_{-1}^{\mu_\star} (c_1-\Omega_1)(1-\mu^2)\,\mathrm{d}\mu}+  \frac{c_2}{\displaystyle \int_{\mu_\star}^1 (c_1-\Omega_1)(1-\mu^2)\,\mathrm{d}\mu}=\frac{2\mathrm{i}\pi c_1}{|\beta_\star'|(1-\mu_\star^2)^2}\, .
\end{equation}
Using the original variables, this equation becomes
\begin{equation}\label{A10}
  \frac{c-\Omega_\star}{\displaystyle \int_{-1}^{\mu_\star} (c-\Omega)(1-\mu^2)\,\mathrm{d}\mu}+  \frac{c-\Omega_\star}{\displaystyle \int_{\mu_\star}^1 (c-\Omega)(1-\mu^2)\,\mathrm{d}\mu}=\frac{2\mathrm{i}\pi c}{|\beta_\star'|(1-\mu_\star^2)^2}\,,
\end{equation}
which we may now compare to (\ref{4.21}). The only difference is that the $\Omega_\star$ on the right-hand side has now been replaced by $c$. However, according (\ref{A7}), $c$ and $\Omega_\star$ are the same up to order $O(\varepsilon^2)$, so (\ref{A10}) and (\ref{4.21}) are equivalent in the limit of small $\varepsilon$, and we may use the latter as the uniform expression.

\begin{figure}
  \centering
  \includegraphics[width=0.49\linewidth]{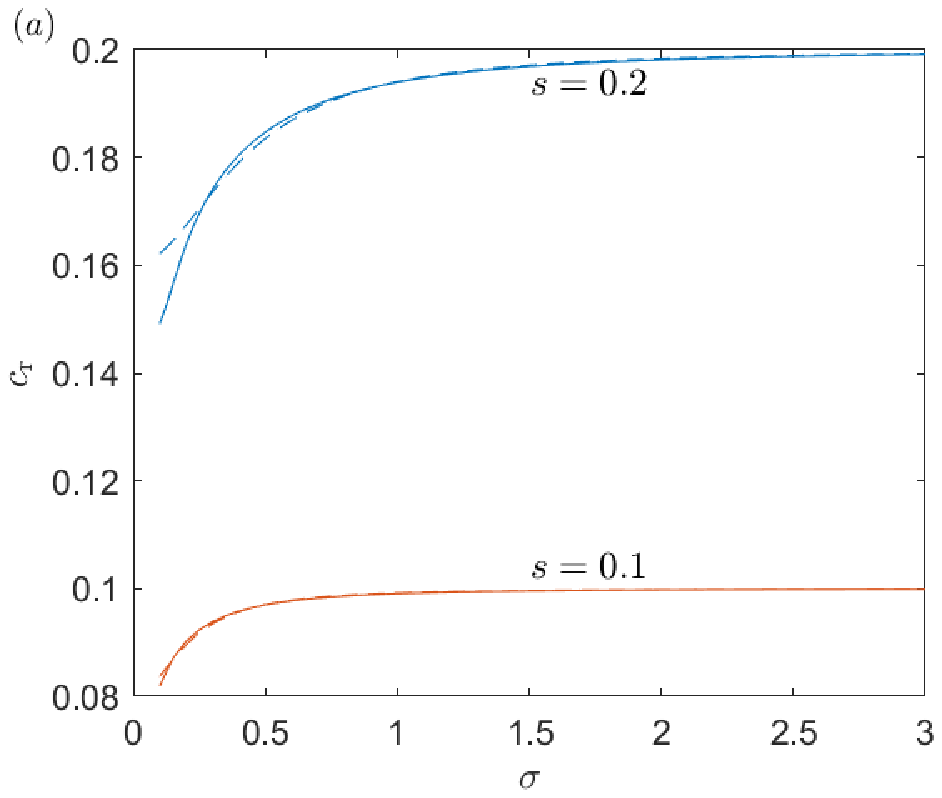}
  \includegraphics[width=0.49\linewidth]{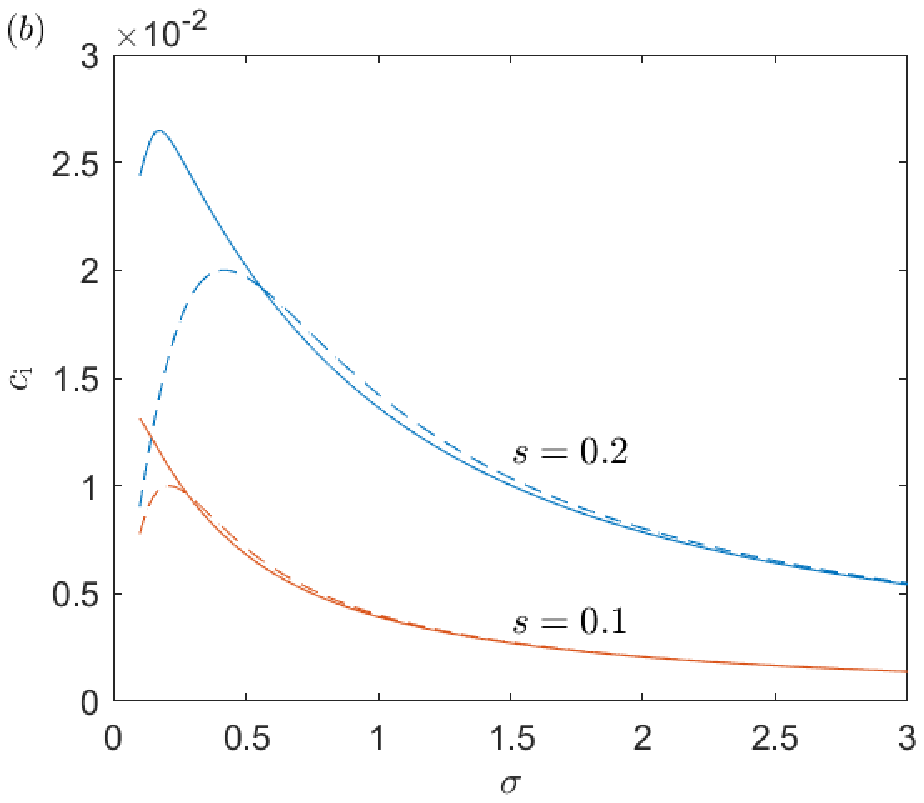}
  \caption{ Eigenvalue $c=c_\mathrm{r}+\mathrm{i}c_\mathrm{i}$ versus $\sigma$ for the basic state $\Omega=s(1-\mu^2)$, $\beta=\sigma\mu$ with $s=0.1$ and $s=0.2$. Solid lines represent numerical solutions, and dashed lines represent the asymptotic solution (\ref{4.30}) with $r=s$.}\label{F7}
\end{figure}

To demonstrate the accuracy of the asymptotic solution we consider the standard flow  (\ref{4.30a}) with $r=s$:
\begin{equation}\label{A11}
  \Omega=s-s\mu^2,\quad \beta=\sigma \mu.
\end{equation}
In this case,  there is no longer a solid body rotation to leading order in $\Omega$,  but our analysis indicates that the solution  (\ref{4.30}) with  $r=s$ is  still valid  when $s$ is small.   The comparison between (\ref{4.30}) with $r=s$ and the numerical solution is plotted in figure~\ref{F7}. We see that the asymptotic solution is very precise for most parameters. It only fails when $\sigma$ becomes small, and in this case the assumption that $\Omega \ll \beta$ is no longer valid.

%
%

%
%
%
%

\section{The conservation of mean toroidal field}
Performing an analysis similar to \S \ref{S5.5} for the mean toroidal field governed by (\ref{2.32b}), we find
\begin{equation}\label{B1}
  \int_0^{\theta_\star-\Delta}\frac{\partial\Delta A}{\partial t}\, \mathrm{d}\theta=-2c_\mathrm{i}\beta_\star'|A_+|^2,\quad \int_{\theta_\star+\Delta}^\pi\frac{\partial \Delta A}{\partial t}\, \mathrm{d}\theta=2c_\mathrm{i}\beta_\star'|A_-|^2,
\end{equation}
and
\begin{equation}\label{B2}
  \int_{\theta_\star-\Delta}^{\theta_\star+\Delta}\frac{\partial\Delta A}{\partial t}\, \mathrm{d}\theta=-2c_\mathrm{i}\beta_\star'\left\{\frac{\pi^2|\alpha_1|^2}{|\Omega_{1\star}-c_1|^2\beta_\star'^2}+2\mathrm{Im}\left[\frac{\pi\alpha_1\alpha_2^*}{(\Omega_{1\star}-c_1)|\beta_\star'|}\right]\right\}.
\end{equation}
Given the conservation law (\ref{2.34}), the mean-field modification in the critical layer is therefore responsible for the difference between $|A_-|$ and $|A_+|$, i.e. the amplitudes of the tilting modes on the two sides of the critical layer. However, such a difference  is not necessary for the instability, and it is also not easy to determine the sign of (\ref{B2}) without further knowledge of the relation between $\alpha_1$ and $\alpha_2$; we conclude that limited insights can be drawn from this conservation law.


\bibliographystyle{jfm}
\bibliography{jfm-instructions}

\begin{thebibliography}{31}
\expandafter\ifx\csname natexlab\endcsname\relax\def\natexlab#1{#1}\fi
\def\au#1{#1} \def\ed#1{#1} \def\yr#1{#1}\def\at#1{#1}\def\jt#1{\textit{#1}}
  \def\bt#1{#1}\def\bvol#1{\textbf{#1}} \def\vol#1{#1} \def\pg#1{#1}
  \def\publ#1{#1}\def\arxiv#1{#1}\def\org#1{#1}\def\st#1{\textit{#1}}

\bibitem[Bernoff \& Lingevitch(1994)]{Bernoff94}
{\sc \au{Bernoff, A.~J.} \& \au{Lingevitch, J.~F.}} \yr{1994}  \at{{Rapid
  relaxation of an axisymmetric vortex}}.  \jt{Phys. Fluids.}  \bvol{6},
  \pg{3717--3723}.

\bibitem[Briggs {\em et~al.\/}(1970)Briggs, Daugherty \& Levy]{Briggs70}
{\sc \au{Briggs, R.~J.}, \au{Daugherty, J.~D.} \& \au{Levy, R.~H.}} \yr{1970}
  \at{Role of {L}andau damping in crossed-field electron beams and inviscid
  shear flow}.  \jt{Phys. Fluids}  \bvol{13}~(2),  \pg{421--432}.

\bibitem[Brun \& Browning(2017)]{Brun17}
{\sc \au{Brun, A.~S.} \& \au{Browning, M.~K.}} \yr{2017}  \at{Magnetism, dynamo
  action and the solar-stellar connection}.  \jt{Living Rev. Sol. Phys.}
  \bvol{14}~(1),  \pg{1--133}.

\bibitem[{Cally}(2000)]{Cally00}
{\sc \au{{Cally}, P.~S.}} \yr{2000}  \at{A sufficient condition for instability
  in a sheared incompressible magnetofluid}.  \jt{Sol. Phys.}  \bvol{194}~(2),
  \pg{189--196}.

\bibitem[Cally(2001)]{Cally01}
{\sc \au{Cally, P.~S.}} \yr{2001}  \at{Nonlinear evolution of 2{D} tachocline
  instabilities}.  \jt{Sol. Phys.}  \bvol{199}~(2),  \pg{231--249}.

\bibitem[Cally {\em et~al.\/}(2003)Cally, Dikpati \& Gilman]{Cally03}
{\sc \au{Cally, P.~S.}, \au{Dikpati, M.} \& \au{Gilman, P.~A}} \yr{2003}
  \at{Clamshell and tipping instabilities in a two-dimensional
  magnetohydrodynamic tachocline}.  \jt{Astrophys. J.}  \bvol{582}~(2),
  \pg{1190--1205}.

\bibitem[Chandra(1973)]{Chandra73}
{\sc \au{Chandra, K.}} \yr{1973}  \at{Hydromagnetic stability of plane
  heterogeneous shear flow}.  \jt{J. Phys. Soc. Japan}  \bvol{34}~(2),
  \pg{539--542}.

\bibitem[Charbonneau(2014)]{Charbonneau14}
{\sc \au{Charbonneau, P.}} \yr{2014}  \at{Solar dynamo theory}.  \jt{Annu. Rev.
  Astron. Astrophys.}  \bvol{52}~(1),  \pg{251--290}.

\bibitem[Deguchi(2021)]{Deguchi21}
{\sc \au{Deguchi, K.}} \yr{2021}  \at{Eigenvalue bounds for compressible
  stratified magnetoshear flows varying in two transverse directions}.  \jt{J.
  Fluid Mech.}  \bvol{920}.

\bibitem[Dikpati \& Gilman(1999)]{Dikpati99}
{\sc \au{Dikpati, M.} \& \au{Gilman, P.~A.}} \yr{1999}  \at{Joint instability
  of latitudinal differential rotation and concentrated toroidal fields below
  the solar convection zone}.  \jt{Astrophys. J.}  \bvol{512}~(1),
  \pg{417--441}.

\bibitem[Drazin \& Reid(1982)]{Drazin82}
{\sc \au{Drazin, P.~G.} \& \au{Reid, W.~H.}} \yr{1982} {\em Hydrodynamic
  stability\/}.  \publ{Cambridge {U}niversity {P}ress}.

\bibitem[Gilman(1967)]{Gilman67}
{\sc \au{Gilman, P.~A.}} \yr{1967}  \at{Stability of baroclinic flows in a
  zonal magnetic field: part {I}}.  \jt{J. Atmos. Sci.}  \bvol{24}~(2),
  \pg{101--118}.

\bibitem[Gilman \& Dikpati(2000)]{Gilman00b}
{\sc \au{Gilman, P.~A.} \& \au{Dikpati, M.}} \yr{2000}  \at{Joint instability
  of latitudinal differential rotation and concentrated toroidal fields below
  the solar convection zone. {II}. instability of narrow bands at all
  latitudes}.  \jt{Astrophys. J.}  \bvol{528}~(1),  \pg{552}.

\bibitem[Gilman \& Dikpati(2002)]{Gilman02}
{\sc \au{Gilman, P.~A.} \& \au{Dikpati, M.}} \yr{2002}  \at{Analysis of
  instability of latitudinal differential rotation and toroidal field in the
  solar tachocline using a magnetohydrodynamic shallow-water model. {I}.
  {I}nstability for broad toroidal field profiles}.  \jt{Astrophys. J.}
  \bvol{576}~(2),  \pg{1031--1047}.

\bibitem[Gilman \& Fox(1997)]{Gilman97}
{\sc \au{Gilman, P.~A.} \& \au{Fox, P.~A.}} \yr{1997}  \at{Joint instability of
  latitudinal differential rotation and toroidal magnetic fields below the
  solar convection zone}.  \jt{Astrophys. J.}  \bvol{484}~(1),  \pg{439--454}.

\bibitem[Gilman \& Fox(1999)]{Gilman99}
{\sc \au{Gilman, P.~A.} \& \au{Fox, P.~A.}} \yr{1999}  \at{Joint instability of
  latitudinal differential rotation and toroidal magnetic fields below the
  solar convection zone. {II} {Instability} for toroidal fields that have a
  node between the equator and pole}.  \jt{Astrophys. J.}  \bvol{510}~(2),
  \pg{1018--1044}.

\bibitem[Gough(2007)]{Gough07}
{\sc \au{Gough, D.~O.}} \yr{2007}  \at{{An introduction to the solar
  tachocline}}.  \bt{In {\em The solar tachocline\/} (ed. \ed{D.~W. Hughes,
  R.~Rosner \& N.~O. Weiss})},  \pg{pp. 1--30}.  \publ{Cambridge University
  Press}.

\bibitem[Howard(1961)]{Howard61}
{\sc \au{Howard, L.~N.}} \yr{1961}  \at{Note on a paper of {J}ohn {W}.
  {M}iles}.  \jt{J. Fluid Mech.}  \bvol{10}~(4),  \pg{509--512}.

\bibitem[Howard \& Gupta(1962)]{Howard62}
{\sc \au{Howard, L.~N.} \& \au{Gupta, A.~S.}} \yr{1962}  \at{On the
  hydrodynamic and hydromagnetic stability of swirling flows}.  \jt{J. Fluid
  Mech.}  \bvol{14}~(3),  \pg{463--476}.

\bibitem[Hughes \& Tobias(2001)]{Hughes01}
{\sc \au{Hughes, D.~W.} \& \au{Tobias, S.~M.}} \yr{2001}  \at{On the
  instability of magnetohydrodynamic shear flows}.  \jt{Proc. R. Soc. A}
  \bvol{457}~(2010),  \pg{1365--1384}.

\bibitem[M{\'a}rquez-Artavia {\em et~al.\/}(2017)M{\'a}rquez-Artavia, Jones \&
  Tobias]{Marquez17}
{\sc \au{M{\'a}rquez-Artavia, X.}, \au{Jones, C.~A.} \& \au{Tobias, S.~M.}}
  \yr{2017}  \at{Rotating magnetic shallow water waves and instabilities in a
  sphere}.  \jt{Geophys. Astrophys. Fluid Dyn.}  \bvol{111}~(4),
  \pg{282--322}.

\bibitem[Miesch(2007)]{Miesch07sustained}
{\sc \au{Miesch, M.~S.}} \yr{2007}  \at{Sustained magnetoshear instabilities in
  the solar tachocline}.  \jt{Astrophys. J.}  \bvol{658}~(2),  \pg{L131}.

\bibitem[Miesch {\em et~al.\/}(2007)Miesch, Gilman \& Dikpati]{Miesch07}
{\sc \au{Miesch, M.~S.}, \au{Gilman, P.~A.} \& \au{Dikpati, M.}} \yr{2007}
  \at{Nonlinear evolution of global magnetoshear instabilities in a
  three-dimensional thin-shell model of the solar tachocline}.  \jt{Astrophys.
  J., Suppl. Ser.}  \bvol{168}~(2),  \pg{337}.

\bibitem[Newton \& Nunn(1951)]{Newton51}
{\sc \au{Newton, H.~W.} \& \au{Nunn, M.~L.}} \yr{1951}  \at{The {S}un's
  rotation derived from sunspots 1934--1944 and additional results}.  \jt{Mon.
  Notices Royal Astron. Soc.}  \bvol{111}~(4),  \pg{413--421}.

\bibitem[Riedinger \& Gilbert(2014)]{Riedinger}
{\sc \au{Riedinger, X.} \& \au{Gilbert, A.~D.}} \yr{2014}  \at{Critical layer
  and radiative instabilities in shallow-water shear flows}.  \jt{J. Fluid
  Mech.}  \bvol{751},  \pg{539--569}.

\bibitem[Sasaki {\em et~al.\/}(2012)Sasaki, Takehiro \& Yamada]{Sasaki12}
{\sc \au{Sasaki, E.}, \au{Takehiro, S.} \& \au{Yamada, M.}} \yr{2012}  \at{A
  note on the stability of inviscid zonal jet flows on a rotating sphere}.
  \jt{J. Fluid Mech.}  \bvol{710},  \pg{154--165}.

\bibitem[Sharif \& Jones(2005)]{Sharif05}
{\sc \au{Sharif, B.~W.} \& \au{Jones, C.~A.}} \yr{2005}  \at{Rotational and
  magnetic instability in the diffusive tachocline}.  \jt{Geophys. Astrophys.
  Fluid Dyn.}  \bvol{99}~(6),  \pg{493--511}.

\bibitem[Thuburn \& Haynes(1996)]{Thuburn96}
{\sc \au{Thuburn, J.} \& \au{Haynes, P.~H.}} \yr{1996}  \at{Bounds on the
  growth rate and phase velocity of instabilities in non-divergent barotropic
  flow on a sphere: A semicircle theorem}.  \jt{Q. J. R. Meteorol. Soc.}
  \bvol{122}~(531),  \pg{779--787}.

\bibitem[Wang \& Balmforth(2018)]{Wang2018}
{\sc \au{Wang, C.} \& \au{Balmforth, N.~J.}} \yr{2018}  \at{Strato-rotational
  instability without resonance}.  \jt{J. Fluid Mech.}  \bvol{846},
  \pg{815--833}.

\bibitem[Wang {\em et~al.\/}(2022)Wang, Gilbert \& Mason]{Wang22}
{\sc \au{Wang, C.}, \au{Gilbert, A.~D.} \& \au{Mason, J.}} \yr{2022}
  \at{Critical-layer instability of shallow water magnetohydrodynamic shear
  flows}.  \jt{J. Fluid Mech.}  \bvol{943},  \pg{A12}.

\bibitem[Watson(1981)]{Watson81}
{\sc \au{Watson, M.}} \yr{1981}  \at{Shear instability of differential rotation
  in stars}.  \jt{Geophys. Astrophys. Fluid Dyn.}  \bvol{16}~(1),
  \pg{285--298}.

\end{thebibliography}


\begin{thebibliography}{41}
\expandafter\ifx\csname natexlab\endcsname\relax\def\natexlab#1{#1}\fi
\def\au#1{#1} \def\ed#1{#1} \def\yr#1{#1}\def\at#1{#1}\def\jt#1{\textit{#1}}
  \def\bt#1{#1}\def\bvol#1{\textbf{#1}} \def\vol#1{#1} \def\pg#1{#1}
  \def\publ#1{#1}\def\arxiv#1{#1}\def\org#1{#1}\def\st#1{\textit{#1}}

\bibitem[Arscott {\em et~al.\/}(1995)Arscott, Slavyanov, Schmidt, Wolf, Maroni
  \& Duval]{Heun}
{\sc \au{Arscott, F.~M.}, \au{Slavyanov, S.~Y.}, \au{Schmidt, D.}, \au{Wolf,
  G.}, \au{Maroni, P.} \& \au{Duval, A.}} \yr{1995} {\em Heun's differential
  equations\/}.  \publ{Clarendon Press}.

\bibitem[Balbus \& Hawley(1991)]{Balbus91}
{\sc \au{Balbus, S.~A.} \& \au{Hawley, J.~F.}} \yr{1991}  \at{A powerful local
  shear instability in weakly magnetized disks. {I}-{L}inear analysis.}
  \jt{Astrophys. J.}  \bvol{376},  \pg{214--222}.

\bibitem[Balmforth(1999)]{Balmforth}
{\sc \au{Balmforth, N.~J.}} \yr{1999}  \at{Shear instability in shallow water}.
   \jt{J. Fluid Mech.}  \bvol{387},  \pg{97--127}.

\bibitem[Balmforth {\em et~al.\/}(1997)Balmforth, del Castillo-Negrete \&
  Young]{balmforth97}
{\sc \au{Balmforth, N.~J.}, \au{del Castillo-Negrete, D.} \& \au{Young, W.~R.}}
  \yr{1997}  \at{Dynamics of vorticity defects in shear}.  \jt{J. Fluid Mech.}
  \bvol{333},  \pg{197--230}.

\bibitem[Balmforth {\em et~al.\/}(2001)Balmforth, Llewellyn~Smith \&
  Young]{Balmforth_01}
{\sc \au{Balmforth, N.~J.}, \au{Llewellyn~Smith, S.~G.} \& \au{Young, W.~R.}}
  \yr{2001}  \at{Disturbing vortices}.  \jt{J. Fluid Mech.}  \bvol{426},
  \pg{95--133}.

\bibitem[Bender \& Orszag(2013)]{Bender}
{\sc \au{Bender, C.~M.} \& \au{Orszag, S.~A.}} \yr{2013} {\em Advanced
  mathematical methods for scientists and engineers {I}: Asymptotic methods and
  perturbation theory\/}.  \publ{Springer Science \& Business Media}.

\bibitem[Blumen {\em et~al.\/}(1975)Blumen, Drazin \& Billings]{Blumen75}
{\sc \au{Blumen, W.}, \au{Drazin, P.~G.} \& \au{Billings, D.~F.}} \yr{1975}
  \at{Shear layer instability of an inviscid compressible fluid. {P}art 2}.
  \jt{J. Fluid Mech.}  \bvol{71}~(2),  \pg{305--316}.

\bibitem[Booker \& Bretherton(1967)]{Booker}
{\sc \au{Booker, J.~R.} \& \au{Bretherton, F.~P.}} \yr{1967}  \at{{The critical
  layer for internal gravity waves in a shear flow}}.  \jt{J. Fluid Mech.}
  \bvol{27},  \pg{513--539}.

\bibitem[Bretherton(1966)]{Bretherton66critical}
{\sc \au{Bretherton, F.~P.}} \yr{1966}  \at{Critical layer instability in
  baroclinic flows}.  \jt{Q. J. R. Meteorol. Soc.}  \bvol{92}~(393),
  \pg{325--334}.

\bibitem[Briggs {\em et~al.\/}(1970)Briggs, Daugherty \& Levy]{Briggs70}
{\sc \au{Briggs, R.~J.}, \au{Daugherty, J.~D.} \& \au{Levy, R.~H.}} \yr{1970}
  \at{Role of {L}andau damping in crossed-field electron beams and inviscid
  shear flow}.  \jt{Phys. Fluids}  \bvol{13}~(2),  \pg{421--432}.

\bibitem[B\"{u}hler(2014)]{Buhler14}
{\sc \au{B\"{u}hler, O.}} \yr{2014} {\em Waves and mean flows\/}, 2nd edn.
  \publ{{Cambridge University Press}}.

\bibitem[Chandra(1973)]{Chandra73}
{\sc \au{Chandra, K.}} \yr{1973}  \at{Hydromagnetic stability of plane
  heterogeneous shear flow}.  \jt{J. Phys. Soc. Japan}  \bvol{34}~(2),
  \pg{539--542}.

\bibitem[Chandrasekhar(1961)]{Chandrasekharbook}
{\sc \au{Chandrasekhar, S.}} \yr{1961} {\em Hydrodynamic and hydromagnetic
  stability\/}.  \publ{Courier Corporation}.

\bibitem[Chen \& Hasegawa(1974)]{Chen1974pulsation}
{\sc \au{Chen, L.} \& \au{Hasegawa, A.}} \yr{1974}  \at{A theory of long-period
  magnetic pulsations: 1. {S}teady state excitation of field line resonance}.
  \jt{J. Geophys. Res.}  \bvol{79}~(7),  \pg{1024--1032}.

\bibitem[Chen \& Morrison(1991)]{Morrison91}
{\sc \au{Chen, X.~L.} \& \au{Morrison, P.~J.}} \yr{1991}  \at{A sufficient
  condition for the ideal instability of shear flow with parallel magnetic
  field}.  \jt{Phys. Fluids B}  \bvol{3}~(4),  \pg{863--865}.

\bibitem[Dellar(2002)]{Dellar}
{\sc \au{Dellar, P.~J.}} \yr{2002}  \at{{Hamiltonian and symmetric hyperbolic
  structures of shallow water magnetohydrodynamics}}.  \jt{Phys. Plasmas}
  \bvol{9},  \pg{1130--1136}.

\bibitem[Dritschel {\em et~al.\/}(2018)Dritschel, Diamond \&
  Tobias]{Dritschel18}
{\sc \au{Dritschel, D.~G.}, \au{Diamond, P.~H.} \& \au{Tobias, S.~M.}}
  \yr{2018}  \at{Circulation conservation and vortex breakup in
  magnetohydrodynamics at low magnetic prandtl number}.  \jt{J. Fluid Mech.}
  \bvol{857},  \pg{38--60}.

\bibitem[Gilman(2000)]{Gilman2000}
{\sc \au{Gilman, P.~A.}} \yr{2000}  \at{Magnetohydrodynamic ``shallow water''
  equations for the solar tachocline}.  \jt{Astrophys. J. Lett.}
  \bvol{544}~(1),  \pg{L79--L82}.

\bibitem[Gilman \& Dikpati(2002)]{Gilman02}
{\sc \au{Gilman, P.~A.} \& \au{Dikpati, M.}} \yr{2002}  \at{Analysis of
  instability of latitudinal differential rotation and toroidal field in the
  solar tachocline using a magnetohydrodynamic shallow-water model. {I}.
  {I}nstability for broad toroidal field profiles}.  \jt{Astrophys. J.}
  \bvol{576}~(2),  \pg{1031--1047}.

\bibitem[Gilman \& Fox(1997)]{Gilman97}
{\sc \au{Gilman, P.~A.} \& \au{Fox, P.~A.}} \yr{1997}  \at{Joint instability of
  latitudinal differential rotation and toroidal magnetic fields below the
  solar convection zone}.  \jt{Astrophys. J.}  \bvol{484}~(1),  \pg{439--454}.

\bibitem[Hayashi \& Young(1987)]{hayashi}
{\sc \au{Hayashi, Y.-Y.} \& \au{Young, W.~R.}} \yr{1987}  \at{{Stable and
  unstable shear modes of rotating parallel flows in shallow water}}.  \jt{J.
  Fluid Mech.}  \bvol{184},  \pg{477--504}.

\bibitem[Heifetz {\em et~al.\/}(2015)Heifetz, Mak, Nycander \&
  Umurhan]{Heifetz15}
{\sc \au{Heifetz, E.}, \au{Mak, J.}, \au{Nycander, J.} \& \au{Umurhan, O.~M}}
  \yr{2015}  \at{Interacting vorticity waves as an instability mechanism for
  magnetohydrodynamic shear instabilities}.  \jt{J. Fluid Mech.}  \bvol{767},
  \pg{199--225}.

\bibitem[Hinch(1991)]{Hinch}
{\sc \au{Hinch, E.~J.}} \yr{1991} {\em Perturbation methods\/}.
  \publ{Cambridge University Press}.

\bibitem[Howard(1961)]{Howard61}
{\sc \au{Howard, L.~N.}} \yr{1961}  \at{Note on a paper of {J}ohn {W}.
  {M}iles}.  \jt{J. Fluid Mech.}  \bvol{10}~(4),  \pg{509--512}.

\bibitem[Hughes \& Tobias(2001)]{hughes01}
{\sc \au{Hughes, D.~W.} \& \au{Tobias, S.~M.}} \yr{2001}  \at{On the
  instability of magnetohydrodynamic shear flows}.  \jt{Proc. R. Soc. A}
  \bvol{457}~(2010),  \pg{1365--1384}.

\bibitem[Kent(1968)]{Kent}
{\sc \au{Kent, A.}} \yr{1968}  \at{Stability of laminar magnetofluid flow along
  a parallel magnetic field}.  \jt{J. Plasma Phys.}  \bvol{2}~(4),
  \pg{543--556}.

\bibitem[Killworth \& McIntyre(1985)]{McIntyre85}
{\sc \au{Killworth, P.~D.} \& \au{McIntyre, M.~E.}} \yr{1985}  \at{Do
  {Rossby}-wave critical layers absorb, reflect, or over-reflect?}  \jt{J.
  Fluid Mech.}  \bvol{161},  \pg{449--492}.

\bibitem[Lecoanet {\em et~al.\/}(2010)Lecoanet, Zweibel, Townsend \&
  Huang]{Lecoanet10}
{\sc \au{Lecoanet, D.}, \au{Zweibel, E.~G.}, \au{Townsend, R. H.~D.} \&
  \au{Huang, Y.-M.}} \yr{2010}  \at{Violation of {R}ichardson's criterion via
  introduction of a magnetic field}.  \jt{Astrophys. J.}  \bvol{712}~(2),
  \pg{1116--1128}.

\bibitem[Mak {\em et~al.\/}(2016)Mak, Griffiths \& Hughes]{Mak16}
{\sc \au{Mak, J.}, \au{Griffiths, S.~D.} \& \au{Hughes, D.~W.}} \yr{2016}
  \at{Shear flow instabilities in shallow-water magnetohydrodynamics}.  \jt{J.
  Fluid Mech.}  \bvol{788},  \pg{767--796}.

\bibitem[Miles(1957)]{Miles57}
{\sc \au{Miles, J.~W.}} \yr{1957}  \at{On the generation of surface waves by
  shear flows}.  \jt{J. Fluid Mech.}  \bvol{3},  \pg{185--204}.

\bibitem[Mok \& Einaudi(1985)]{Mok85}
{\sc \au{Mok, Y.} \& \au{Einaudi, G.}} \yr{1985}  \at{Resistive decay of
  {A}lfv{\'e}n waves in a non-uniform plasma}.  \jt{J. Plasma Phys.}
  \bvol{33}~(2),  \pg{199--208}.

\bibitem[Riedinger \& Gilbert(2014)]{Riedinger}
{\sc \au{Riedinger, X.} \& \au{Gilbert, A.~D.}} \yr{2014}  \at{Critical layer
  and radiative instabilities in shallow-water shear flows}.  \jt{J. Fluid
  Mech.}  \bvol{751},  \pg{539--569}.

\bibitem[Sakurai {\em et~al.\/}(1991)Sakurai, Goossens \&
  Hollweg]{Sakurai1991resonant}
{\sc \au{Sakurai, T.}, \au{Goossens, M.} \& \au{Hollweg, J.~V.}} \yr{1991}
  \at{Resonant behaviour of magnetohydrodynamic waves on magnetic flux tubes
  {II}. {A}bsorption of sound waves by sunspots}.  \jt{Sol. Phys.}  \bvol{133},
   \pg{247--262}.

\bibitem[Satomura(1981)]{Satomura81}
{\sc \au{Satomura, T.}} \yr{1981}  \at{An investigation of shear instability in
  a shallow water}.  \jt{J. Meteor. Soc. Japan Ser. II}  \bvol{59}~(1),
  \pg{148--167}.

\bibitem[Shukhman(1998)]{Shukhman98}
{\sc \au{Shukhman, I.~G.}} \yr{1998}  \at{Nonlinear evolution of a weakly
  unstable wave in a free shear flow with a weak parallel magnetic field}.
  \jt{J. Fluid Mech.}  \bvol{369},  \pg{217--252}.

\bibitem[Stern(1963)]{Stern63}
{\sc \au{Stern, M.~E.}} \yr{1963}  \at{Joint instability of hydromagnetic
  fields which are separately stable}.  \jt{Phys. Fluids}  \bvol{6}~(5),
  \pg{636--642}.

\bibitem[Tatsuno \& Dorland(2006)]{Tatsuno06}
{\sc \au{Tatsuno, T.} \& \au{Dorland, W.}} \yr{2006}  \at{Magneto-flow
  instability in symmetric field profiles}.  \jt{Phys. Plasmas}  \bvol{13}~(9),
   \pg{092107}.

\bibitem[Turner \& Gilbert(2007)]{Turner07}
{\sc \au{Turner, M.~R.} \& \au{Gilbert, A.~D.}} \yr{2007}  \at{Linear and
  nonlinear decay of cat's eyes in two-dimensional vortices, and the link to
  {L}andau poles}.  \jt{J. Fluid Mech.}  \bvol{593},  \pg{255--279}.

\bibitem[Vekstein(1998)]{Vekstein98}
{\sc \au{Vekstein, G.~E.}} \yr{1998}  \at{Landau resonance mechanism for plasma
  and wind-generated water waves}.  \jt{Am. J. Phys.}  \bvol{66}~(10),
  \pg{886--892}.

\bibitem[Wang \& Balmforth(2018)]{Wang2018}
{\sc \au{Wang, C.} \& \au{Balmforth, N.~J.}} \yr{2018}  \at{Strato-rotational
  instability without resonance}.  \jt{J. Fluid Mech.}  \bvol{846},
  \pg{815--833}.

\bibitem[Wang \& Balmforth(2021)]{Wang21}
{\sc \au{Wang, C.} \& \au{Balmforth, N.~J.}} \yr{2021}  \at{Nonlinear dynamics
  of forced baroclinic critical layers {II}}.  \jt{J. Fluid Mech.}  \bvol{917}.

\end{thebibliography}

\end{document}